\documentclass[10pt,conference,final]{IEEEtran}

\usepackage{graphicx}
\usepackage[utf8]{inputenc}
\usepackage[T1]{fontenc}
\usepackage[hyphens]{url}
\usepackage{clrscode}
\usepackage{xcolor}
\usepackage{paralist}

 \usepackage{wasysym}
\newcommand{\eg}{e.\,g.,\ } % note the trailing comma (recommended by http://grammar.quickanddirtytips.com/ie-eg-oh-my.aspx )

\definecolor{thomas}{RGB}{0,100,0}

\definecolor{thorsten}{RGB}{0,0,255}

\begin{document}

% \title{The Impact of Comprehensible Audience Selection Controls on the Exposure of User Data in Facebook}
\title{Private Date Exposure in Facebook and the Impact of Comprehensible Audience Selection Controls}

\author{
  \IEEEauthorblockN{Thomas Paul$^+$, ~ Daniel Puscher$^+$, ~ Thorsten Strufe$^*$}
  \IEEEauthorblockA{TU Darmstadt$^+$ ~ and ~ TU Dresden$^*$\\
thomas.paul@cs.tu-darmstadt.de, uni@daniel-puscher.de, thorsten.strufe@tu-dresden.de
   } 
}

\maketitle

\begin{abstract} 

Privacy in Online Social Networks (OSNs) evolved from a niche topic to a broadly discussed issue in a wide variety of media. Nevertheless, OSNs drastically increase the amount of information that can be found about individuals on the web. To estimate the dimension of data leakage in OSNs, we measure the real exposure of user content of 4,182 Facebook users from 102 countries in the most popular OSN, Facebook. We further quantify the impact of a comprehensible privacy control interface that has been shown to extremely decrease configuration efforts as well as misconfiguration in audience selection. 

Our study highlights the importance of usable security. (i) The total amount of content that is visible to Facebook users does not dramatically decrease by simplifying the audience selection interface, but the composition of the visible content changes. (ii) Which information is uploaded to Facebook as well as which information is shared with whom strongly depends on the user's country of origin.   

\end{abstract}

\begin{IEEEkeywords}
% Are NOT: Peer-To-Peer, Anonymity, Privacy.
% BUT TAKEN FROM THIS LIST: 
% http://www.ieee.org/organizations/pubs/ani_prod/keywrd98.txt
Social Networks, Privacy Control, Facebook

\end{IEEEkeywords}

\maketitle

\section{Introduction}

Online Social Networks (OSNs), such as Facebook
% Xing\footnote{www.xing.de} 
or google+, have about one billion users\footnote{http://allfacebook.de/userdata/, Accessed 2015-03-06} in 2015. OSNs allow their users to create and maintain a personal user profile and connect this profile with others by declaring friendship relations. Amongst communication functionalities, sharing content and personal information is the core of OSN sites. Content sharing serves communication and self-expression needs of OSN users, but raises privacy concerns at the same time. %Users have to balance between those two poles by choosing which information they want to share with whom.

There is an ongoing discussion about how to handle those privacy concerns. The CEOs of Google and Facebook argue that we live in a post-privacy world \cite{gogleprivacy,facebookprivacy}. We shall accept the fact that there is no privacy anymore and adapt ourselves to the new situation. On the other side of the discussion spectrum, privacy advocates fear oversharing of content \cite{liu2011analyzing} to avoid undesired effects such as that employers are accessing private information to draw undesired conclusions. In spite of this discussion, the real privacy preferences of the social networking community are still not entirely known. 

\begin{figure}[h]
\centering
\includegraphics[width=0.477\textwidth]{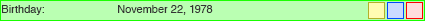}
\caption{Example for an attribute's privacy setting (User's Birthday)}
\label{fig:Example}
\end{figure}

Studying the actual privacy settings of Facebook users (e.g. \cite{krishnamurthy2008characterizing}) does also not tell the whole story about content sharing and privacy preferences, since users are commonly unable to select the desired audience \cite{liu2011analyzing,Madejski}. We thus developed a color-based interface to simplify the audience selection for user content in Facebook (Figure \ref{fig:Example}; detailed description in Section \ref{subsec:fpw}). This interface is shown to drastically decrease both the effort and the error probability when handling privacy settings \cite{paul2012c4ps}. %To help users to meet their  sharing interests, w

This color coding-based interface is published in the shape of an add-on (plug-in) for the Firefox and Chrome browsers and made available to the public over various channels on the web. %to allow the public to benefit from our research. 
This plug-in is called Facebook Privacy Watcher (FPW).
%   The plug-in was published at our homepage\footnote{\url{http://www.daniel-puscher.de/fpw}} and different mirrors. 
Several newspapers, blogs and even radio and TV stations reported about it\footnote{\url{http://www.masrawy.com/news/Technology/General/2012/October/31/5420245.aspx}, Accessed 2015-03-06} \footnote{\url{http://www.golem.de/news/facebook-firefox-erweiterung-macht-die-privatsphaere-bunt-1212-96091.html}, Accessed 2015-03-06}. %\footnote{\url{http://www.ffh.de/news-service/magazin/toController/Topic/toAction/show/toId/3371/toTopic/die-facebook-ampel-fuer-sichere-postings.html, Accessed 2015-03-06}}  %\cite{golem,handelsblatt,masrawy,lifehacker,windowsclub}. 
More than 44,000 users downloaded the FPW.

We asked the FPW users to send us anonymized feedback with consent, to improve the plug-in and to evaluate the impact of the plug-in on user's privacy. We received 9,296 feedback responses originated from 102 countries. These responses included the privacy settings of the user profiles and the changes that were made with the help of our plug-in. Furthermore, we received the number of friends, photos, likes, notes and map entries as well as the binary information for each user profile data field (denoted profile field in the remainder) whether it is filled with data or not.

Based on this dataset, we evaluate the real exposure of private user data in Facebook and the content sharing preferences of the FPW users. We evaluate the privacy settings before and after introducing a comprehensible visualization of privacy handle as well as the changes that have been performed. By reason that the results strongly differ with respect to different countries, we also performed evaluations that focus on national differences.
%We further provide insights into profile size  of FPW users. 
Assuming that increasing or decreasing the visibility of parts of the user profiles expresses the desires of users to have more or less privacy, we compared the user profiles of users who use the FPW to achieve more privacy with those who decided to publish more private data.
%We finalize the evaluation with cluster analyses in regard to the change actions which have been performed with the FPW and their 

Our results indicate that we indeed do not live in a post privacy world:
% According to sharing preferences of FPW users, we do not live in a post-privacy world. 
The users intentionally hide content from being publicly accessed and do not accept the default privacy settings even before using our plug-in. With the help of the FPW, %
users hide critical data fields such as friend lists and family member markers but publish birthdays and religious views. %They tend to insignificantly decrease the number of users who can access parts of their profiles in average. Thus, 
The total amount of content which is visible to Facebook users does not dramatically decrease after introducing a comprehensible visualization of privacy controls, but the composition of the visible content changes. 
The content sharing patterns are strongly depending on their country of origin.

% Overview of our solution and first confidence (how do we show that it's good?)

% Our contributions in this paper
Our contributions in this paper are (i) to provide an understanding of the content sharing preferences of FPW users both in general and (ii) with respect to different countries and (iii) to explain and quantify the effect of improved usability of privacy interfaces on privacy settings. We further (iv) depict relations between privacy preferences and profile properties by means of cluster analyses. An important highlight is that we are not limited to public-available data. Due to the FPW feedback data, we can take the user profile owner's point of view on her privacy settings.

% Reader's digest
The remainder of the paper is organized as follows: We discuss the related work in Section \ref{sec:related_work} and provide a detailed data description in Section \ref{sec:data_description}. In Section \ref{sec:privacy_evaluations}, we evaluate the privacy settings of FPW users and the impact of introducing a comprehensible audience selection without mentioning country specific differences. Because of vast differences amongst users from different countries, we provide a deeper analysis of those specifics in Section \ref{sec:country_specifics}. The relation between sharing preferences and quantifiable user profile properties such as the numbers of friends, likes and photos are evaluated in Section \ref{sec:clusteranalysis}.  We summarize our findings and conclude our work in Section \ref{sec:conclusion}.

% 
% We answer the following questions:
% - do our users change the default settings?
% - which impact does our new privacy interface have on the privacy settings?
% - 

\section{Related Work}
\label{sec:related_work}

%Related work in the field of user behavior in OSN can be found in many scientific disciplines like psychology, sociology and computer science. 

Privacy is a topic that is broadly addressed by plenty of publications in computer science. In this section, we discuss works on privacy in OSNs with the focus on user behavior and interface construction in OSNs rather than systems or algorithms. Since we discuss a new privacy settings interface, default privacy settings and privacy awareness in this paper, we particularly focus on papers about privacy by design as well as on papers suggesting interfaces for privacy settings in OSNs.

% 
% \begin{itemize}
% \item Privacy awareness papers
%  \item PrivDef Papers that 
% \item HCI approaches 
% \end{itemize}

Works on privacy by design are built on the assumption that people do not tend to change their privacy settings. Gross and Acquisti state that "We can conclude that only a vanishingly small number of users change the (permissive) default privacy preferences" \cite{Gross:2005:IRP:1102199.1102214}. Based on this logic, the authors suggest to implement default privacy rules that prevent leakage of data. In contrast to this paper, we evaluate how much a better interface helps the users to meet their needs by avoiding misconfiguration, and compare the sharing preferences with respect to the user's country of origin. Furthermore, our results show that more than 59\% of the privacy settings do not stay untouched in case of using our plug-in.

In 2008, Krishnamurthy and Wills \cite{krishnamurthy2008characterizing} examined privacy settings in Facebook, Myspace, Bebo and Twitter based on crawler-gathered data. They discovered that there is some use of privacy settings but there is still a significant portion of users who allow strangers to access private information. They further examined the amount of information which is shared within regional networks and discovered a negative correlation between network size and the amount of shared information. In comparison to \cite{krishnamurthy2008characterizing}, we focus on Facebook, obtain our data directly from the uses, evaluate the impact of our color-based privacy setting interface and get different results regarding the users disposition to change privacy settings.

Stutzman et al. \cite{stutzman2013silent} monitored the public-available data of 5,076 members of the Carnegie Mellon University from 2005 till 2011. They discovered an increasing privacy awareness over time. Johnson et al. \cite{johnson2012facebook} surveyed 260 participants from the United States, recruited via ResearchMatch, by using a Facebook application. They asked questions with the background knowledge which was obtained by reading the participant's Facebook profile via API. Inter alia, they discovered that 94.6\% of their participants denied access to their content by people outside their friend network. Mondal et al. \cite{mondal2014understanding} studied the use of social access control lists (SACLs).  The friend list usage of 1,165 users of tool ``Friendlist Manager'', has been analyzed. They found ``that a surprisingly large fraction (17.6\%) of content is shared with SACLs. However, we also find that the SACL membership shows little correlation with either profile information or social network links; as a result, it is difficult to predict the subset of a user’s friends likely to appear in a SACL.''

Beside the FPW, other approaches to help users to mitigate the misconfiguration exist, too. Lipford et al. \cite{Lipford_face} suggest to allow users take the point of view of the expected audience. PViz \cite{pviz} is a privacy setting approach based on visualizations group visualizations in different granularities. Carminati et al. \cite{Carminati2006} suggest  rule-based privacy settings that define types of relationships and a set of rules which type of relationship is a precondition to access a certain data object. Fang et al. \cite{Fang2010} propose a machine learning based approach which implements a wizard that suggests a set of access rules. The idea is to learn implicit rules which are applied by users to set the visibility of objects. In contrast, our interface allows both: to quickly grasp the visibility of content items based on a color coding and to change those settings with a single click.

Other related work can be found in studies about Facebook user statistics \cite{wolframalpha}, a report\footnote{http://mattmckeon.com/facebook-privacy/, Accessed 2015-03-06} about the evolution of privacy in Facebook and a survey in \cite{dataprivacy} where consumers have been asked which information they consider to be private. \cite{wolframalpha,dataprivacy} also contain cross-country comparison. However, the user statistics do not provide information about privacy settings and the consumer survey does rely on questionnaires without a concrete link to social networks.
% \tp{Maybe add papers that compare privacy needs with respect to countries}

\section{Experimental Setup and Dataset Description}
\label{sec:data_description}

In this section, we specify the setup of our study by describing our ethical considerations, the browser extension FPW which has been used to collect the data and the precise data collection methods. To underline the adequacy of our color-coding audience selection interface to be used in this study, we describe essences on the feedback from study participants.  We further depict which and how much data we were able to collect in this study and describe basic user profile statistics of the participants. The bias as a result of a non-random selection of study participants is also discussed in this section.

\subsection{Ethical Considerations}

We protect the privacy of our study participants! Neither the download logfile which we used to estimate the dissemination of the FPW, nor the feedback answers that we collected are linked to individuals. For this study, we asked the FPW users to send us feedback with consent. We explained the reason for collecting the data and allowed users to access and verify the data before sending it to our server. All feedback responses that we used in this study are anonymized. % as much as possible.
We keep the collected data confidential to protect all study participants from deanonymization attempts and do only publish aggregated data.

\subsection{The Facebook Privacy Watcher}
\label{subsec:fpw}

In Facebook, the user can choose between a number of visibility-levels for each information in her profile, namely: 'Everyone', 'Friends', 'Custom' and 'Only me'. The 'Custom' setting allows users to authorize single friends or groups of friends (\eg 'colleagues' and 'good friends') to access certain information. %Grouping friends leads to more efficient adjusting of privacy settings  than selecting each user per profile attribute individually. This reflects the communities of acquaintances a user is part of in the real world and helps to decide quickly, which group of contacts shall be granted access to a certain profile entry.
In previous work \cite{paul2012c4ps}, a new type of interface has been presented, which is based on a color coding (Figure \ref{fig:Example}). The used colors are guided by the well-known traffic light colors, adding blue to represent custom settings. Results of changes (initiated by clicking at the respective color box) are shown instantly for \emph{direct success control} of each action. We used the following color scheme:

\begin{itemize}
\item \emph{Red:} Visible to nobody
\item \emph{Blue:} Visible to selected friends
\item \emph{Yellow:} Visible to all friends
\item \emph{Green:} Visible to everyone
\end{itemize}

%
% \begin{figure}[h]
% \centering
% \includegraphics[width=0.45\textwidth]{grafiken/new_privacy_settings_buttons.png}
% \caption{Example for an attribute's privacy setting (``Birthday: 22. Nov. 1978)}
% \label{fig:Example}
% \end{figure}

\begin{figure}[ht]
\centering
\includegraphics[width=0.36\textwidth]{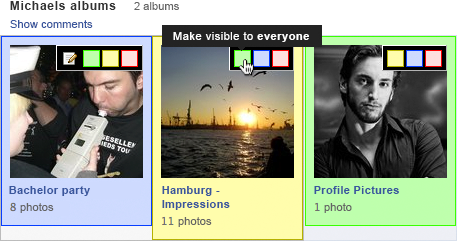}
\caption{Screenshots of photo albums, colorized by the FPW}
\label{fig:screenshot}
\end{figure}

This color scheme is in-line with the sympathy of the majority of the users (Section \ref{subsec:opinions}). Figure \ref{fig:screenshot} shows a screenshot of colorized photo albums as an example for colorized profile fields. Clicking at the colorized boxes changes the privacy settings. A tooltip helps the users to remember the meaning of each color. The color scheme can be adopted to individual user needs. To help color-blind FPW users, we included the possibility to use different stripy patterns instead of colors. The FPW has support for English, German, French, Italian and Arabic.

% Furthermore, we conducted a user study to elaborate whether our new or the original privacy setting interface of Facebook is more intuitive. The results of the user study underlined a clear advantage for the new type of interface. The participants solved all tasks quicker and almost without mistakes.

\subsection{Data Collection}

We gathered data about the FPW from two sources. The first is the download log file at our own server, where the plug-in can be downloaded from. The second source of data is the set of feedback responses which have been sent to us. While the first source gives us insights into the spreading process of the plug-in, the second source allows us to draw a picture of the plug-in usage as well as its impact on privacy settings of the users' profiles.

\subsubsection{Download Log}

% \begin{figure}[ht!]
% \includegraphics[width=0.489\textwidth]{grafiken/feedbacks_per_country.pdf}
% \caption{Feedbacks per country; Top eight countries}
% \label{fig:feedbacks_per_country}
% \end{figure}

Analyzing the download logfile enabled us to understand the time and locality dimensions of the FPW dissemination. We discovered strong peaks subsequently to the moments of publication in different venues as well as that a large user basis is originated in Germany and Egypt. We further discovered a couple of sites, offering to download our plug-in\footnote{\url{http://www.chip.de/downloads/Facebook-Privacy-Watcher-fuer-Firefox_57997141.html, Accessed 2015-03-06}}\footnote{\url{http://www.computerbild.de/download/Facebook-Privacy-Watcher-7834052.html, Accessed 2015-03-06}} %\cite{chip,computerbild,netzwelt,freeware,soft_ware}. 
Thus, we only have an incomplete view on the actual downloads by analyzing our own download log. Some of those alternative download sites publish the number of downloads. Adding the number of downloads from our site to those external download counters, we estimate the total number of download to be higher than 44,800, coming at least from 102 countries. One year after our first FPW publication, 11,000 users are still following every update that we offer.

% Thus, our statistics just includes the downloads from our own site. The download logfile contains information about the User-Agent settings, the IP of the client which downloaded the file, the host name and the timestamp of the download. An external skript added the Country to our data, based on the IP address.

% \begin{center}
% \begin{tabular}{ll}
% Germany & 13694\\
% Egypt & 3609\\
% Saudi Arabia & 413\\
% Austria & 381\\
% Switzerland & 312\\
% United States & 277\\
% United Arab Emirates & 133\\
% United Kingdom & 112\\
% Syrian Arab Republic & 91\\
% Morocco & 79\\
% \end{tabular}
% \end{center}

% \tp{Todo: add a comparison that shows the total number of downloads each country and compares with the number of feedbacks from each country to show the tendency to send feedbacks; Maybe bar chart with two colors (downloads vs feedbacks)}

\subsubsection{User Feedback}
\label{sec:data_feedbacks}

The usual life-cycle of an FPW instance starts with the installation process and resumes with a check of the privacy settings of the own  profile during a few sessions (1-5). The plug-in is sparely used afterwards. We asked our users to provide us feedback after activating the plug-in three times, which usually happened within the first days after installation.

We asked for feedback about both: the general idea of coloring the profile items to simplify the privacy settings and the implementation of our plug-in. Furthermore, we offered two text fields to enter comments and suggestions concerning the idea as well as the implementation.
% Until version 0.69.17 beta, we asked the users to submit the coloring database and the log file of the plug-in for two reasons: some feedbacks contained bug descriptions which could not be solved without detailed information from the log file and the coloring information is a valuable database to investigate the impact of our plug-in.
%
% Since we received a lot of bug notifications without log file and database, we decided in to make sending the database and log file obligatory for sending feedback.
We explicitly informed our users about the exact (anonymized) data that we collected. From 2012-10-15 till 2014-07-07,  we received 9,296 feedback responses from 4,182 users in 102 countries that included coloring and log file information. We received multiple answers from users in Germany. We asked German users twice to give us Feedback: once - in German language - at the time before the FPW was internationally spread and  a second time after introducing multiple language packs. We used the more recent feedbacks to replace older onces in the analysis in case of multiple copies from the same user.

\begin{table}
\small
\centering
\begin{tabular}{rl} \hline
Country & \# Feedback responses \\ \hline
Germany & 7,581 \\
Egypt & 272 \\
Austria & 218 \\
United States & 150 \\
Switzerland & 147 \\
France  & 94 \\
Spain& 72\\
Netherlands & 62\\

\hline
\label{tab:numberoffeedbackspercountry}
\end{tabular}
\caption{The number of Feedback responses that we received from the top eight countries}
\end{table}

We collected the following information from our users:

\begin{compactitem}
 \item a hash value of the Facebook - UIN
 \item the counter (including timestamps), indicating how often the plug-in was activated
 \item the visibility of each profile field before the first usage of our plug-in happened %(in levels 0-5)
 \item the visibility of each profile field after using our plug-in %(in levels 0-5)
 \item the type and visibility of timeline entries
 \item the number of friends
 \item the number of photos and labels
 \item the number of likes
\end{compactitem}

Furthermore, our server, which gathered the feedback data, ran a script to extract the countries from which we received the feedback. %The intention behind storing the profile size metrics (numbers of friends, photos and likes) was to get an idea about who uses our plug-in.

\subsection{Users' Acceptance of the FPW}
\label{subsec:opinions}

It is essential for the success of the study that participants are willing integrate the tool in their normal OSN usage and to use it more than only once. The FPW and the realized user interface hence need to be both: beneficial for the participants and easily usable.
Thus, the first question which we asked our users in the feedback formula was: 'How do you like the idea of colorizing in this plug-in?'. The overwhelming majority rated this idea as 'very good' (65.66\%) or 'good' (32.2\%). Less than one percent rated the idea to be 'medium' (0.98\%), 'bad' (0.46\%) or 'very bad' (0.7\%).

% 0:Very good, 1:Good, 2:Bad, 3:Very bad
% \begin{center}
% \begin{tabular}{lll}
% & Var1 & Freq\\
% 1 & 0 & 66.86\\
% 2 & 1 & 31.8\\
% 3 & 2 & 0.58\\
% 4 & 3 & 0.76\\
% \end{tabular}
% \end{center}

% \subsection{Color Scheme}

Creating the color scheme, we argued in the team which type of color scheme is more intuitive to the users: green, inspired by traffic lights meaning 'go' - corresponding in the color scheme to be visible to everybody or green in the meaning of being safe since the item is not visible to anybody. This question has been asked in the previous user study with 40 participants. 60\% of the participants preferred the green to represent the setting meaning 'visible to everybody'. It roughly meets the results in this study (54.83\% vs. 45.17\%). Please note: The FPW equally offers both color schemes and the users are asked to choose in advance. The setting can later be changed. Color blind people have been offered to choose hachures instead of colors.
%
% Which color scheme is more intuitive?
% 0:Green=Visible for all, 1:Visible only for me
% \begin{center}
% \begin{tabular}{lll}
% & Var1 & Freq\\
% 1 & 0 & 58.67\\
% 2 & 1 & 41.33\\
% \end{tabular}
% \end{center}

% \subsection{Implementation}

The second question that we asked the FPW users was: 'How do you like the implementation of this browser extension?'. The implementation was not rated as good as the idea of using colors for setting privacy (Table \ref{tab:implementation}). Evaluating the comments, we can find the following reasons: %The people who preferred green to represent the safe setting where nobody has access were not satisfied that it was not possible to customize the colors in the first three versions.  However, 
The plug-in did not work from 7th of November 2013, 2:30 am, till 8th of November, 3:30 (am, CET), because of Facebook site changes. During this time, we received most of the negative ratings. Furthermore, we suffered from a bug in the first version that delayed the Facebook usage.

\begin{table}[ht]
\small
\centering
\begin{tabular}{rl} \hline
Rating & Percentage\\ \hline
Very good & 32.34\\
Good & 61.34\\
Medium & 3.44\\
Bad & 1.83\\
Very bad  & 1.05\\ \hline
\end{tabular}
\caption{How do you like the implementation of the FPW?}
\label{tab:implementation}
\end{table}

\subsection{Sample Bias and Basic User Profile Statistics}
\label{subsec:sample_bias}

We recruited our sample (FPW users) via an announcement on our homepage and by sending press releases to specialized press. We then witnessed a viral spreading process based on word-of-mouth advertising. The attention of mass media such as news papers\footnote{\url{http://www.handelsblatt.com/technologie/it-tk/it-internet/facebook-privacy-watcher-im-einsatz-gegen-den-daten-kraken-seite-all/7388782-all.html, Accessed 2015-03-06}}%\cite{handelsblatt}
, radio stations\footnote{\url{http://www.ffh.de/news-service/magazin/toController/Topic/toAction/show/toId/3371/toTopic/die-facebook-ampel-fuer-sichere-postings.html, Accessed 2015-03-06}} and an Egyptian web portal \cite{masrawy} followed afterwards. In spite of the broad audience of the respective media, the set of participants is by no means random. %We also cannot provide
We decided not to collect detailed demographic informations about FPW users, since this would be inappropriate for a tool that has been advertised to support user's privacy.
% to collect data on user demographics supports privacy.
Instead, we provide technical information such as statistics about the user profiles (Table \ref{tab:profile_size}) to allow the sample bias to be appraised:

\begin{table}[ht]
\small
\centering
\begin{tabular}{rllll}
  \hline
 $X$ & $X = 0$ & $ \diameter_{X}$ & $\tilde X$ & $\sigma_{X}$ \\
%   $X$ & $X = 0$ & $\diameter_{X}$ & $\tilde X$ & $\sigma_{X}$ \\
  \hline
Friends & 0\% & 148.75 & 96 & 159.53 \\
Photos & 3.43\% & 181.69 & 32 & 572.62 \\
Labels on photos & 34.45\% & 20.54 & 3 & 64.45 \\
Photo albums & 3.49\% & 10.71 & 7 & 20.05 \\
Locations & 17.07\% & 38.68 & 4 & 101.65 \\
Likes & 10.06\% & 90.04 & 36 & 145.33 \\
Notes & 86.94\% & 1.49 & 0 & 19.5 \\
   \hline
\end{tabular}
\caption{Basic profile statistics: percentage of profiles without any entry in field X ($X = 0$) and the average ($ \diameter_{X}$), median ($\tilde X$) and standard deviation ($\sigma_{X}$) of the number of entries in field X}
\label{tab:profile_size}
\end{table}

\begin{figure}[ht]
\centering
\includegraphics[width=0.4\textwidth]{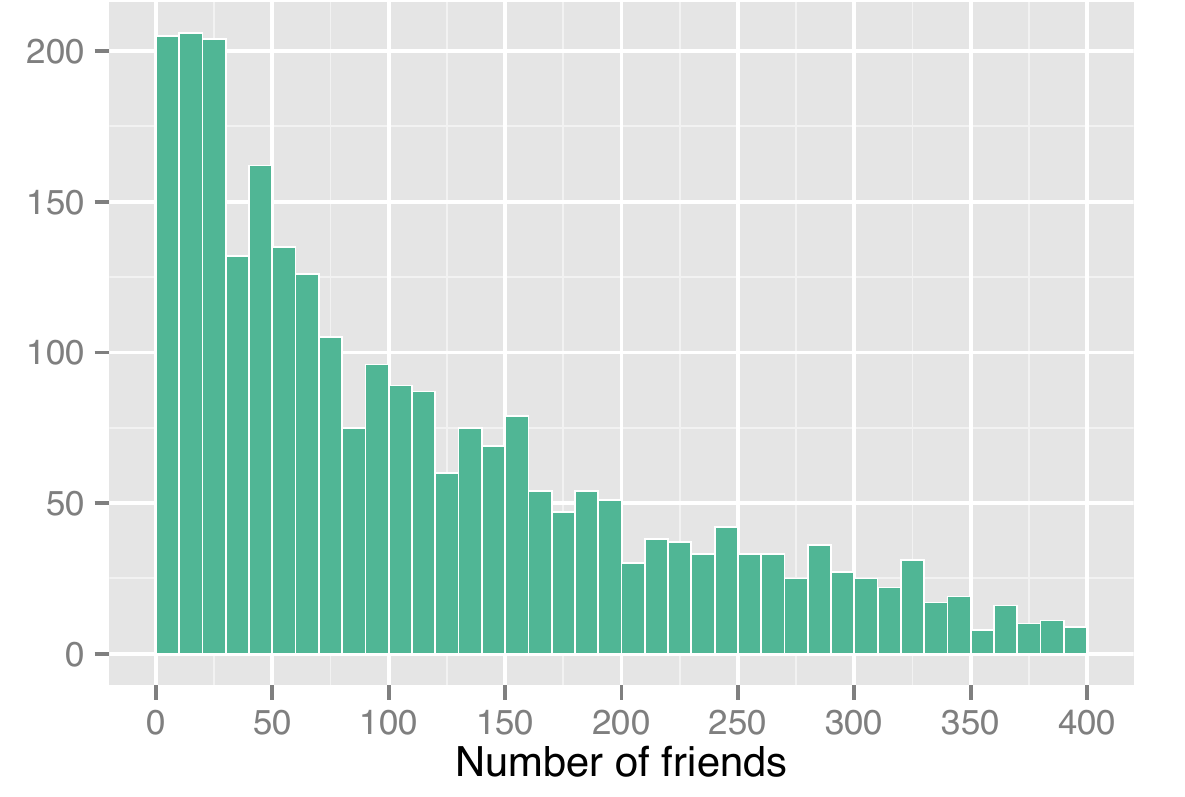}
\caption{Histogram: number of friends}
\label{fig:stats_number_of_friends}
\end{figure}

Our median user has 96 friends, liked 36 pages and shared 32 pictures. Many users have just a few friends (Figure \ref{fig:stats_number_of_friends}) and a few of them have plenty of friends. The degree distribution of the friendship graph as well as the median number of friends is similar to those of the whole Facebook graph \cite{facebook_graph}. We interpret this as an evidence that our FPW users are close to normal with respect to the number of friends. %\tp{ToDo: Compare those numbers with the official numbers from Facebook}

\section{Global Privacy Evaluation}
\label{sec:privacy_evaluations}

In this section, we elaborate which data FPA users upload to Facebook and who is allowed to access it without mentioning cultural differences amongst users from various countries to provide a holistic view. We further quantify the impact of the FPW on the privacy settings and compare the standard privacy settings in Facebook with the actual user decisions to quantify the total demand for modifying the Facebook standard privacy setting to meet users' needs.

Because of the typical life-cycle of the plug-in instances (Section \ref{sec:data_feedbacks}), three data views are available: the privacy settings before using the plug-in, after using the plug-in and the changes that have been made. We avoid the redundancy which would be caused by presenting the three possible points of view. We instead focus on the settings after applying our plug-in and the changes which have been made.

\subsection{Exposure of User Data}

\begin{figure}[ht]
\centering
\includegraphics[width=0.472\textwidth]{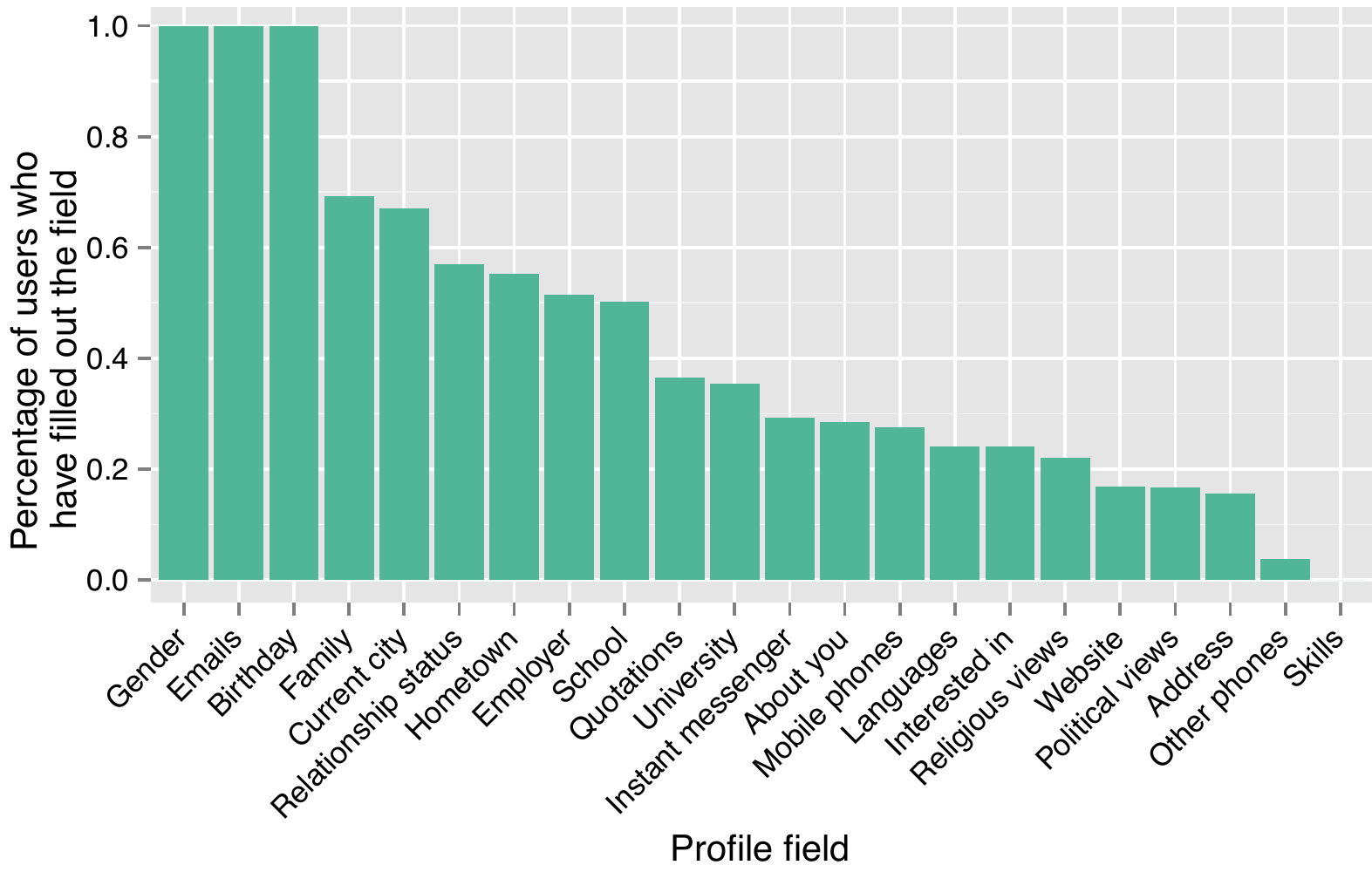}
\caption{Histogram of the ratio, the user profile fields are filled}
\label{fig:filled_profile}
\end{figure}

A Facebook profile can consist of 28 data fields in total. To estimate the potential privacy risk, it is crucial to know which parts of the profile are filled with data and thus potentially exposed to the risk of being accessed by subjects which are not part of the set of desired recipients. The average filling ratio of the profile fields that allow users to select the audience is given in Figure \ref{fig:filled_profile}. 

The profile fields friend list, Timeline entries, photo albums, map entries and notes are lists of items that are technically always available. The number of items included in the users profiles can be found in Tables \ref{tab:cluster_group} and \ref{tab:cluster_country}. Subscriptions are also not included in Figure \ref{fig:filled_profile}. They allow users to follow other users' updates (e.g. news of famous actors) without befriending with them. It is possible to determine the visibility of subscriptions without subscribing anything. %Subscriptions are not included in this Figure, because of the specialty of this feature. 
According to our ethical considerations, we only store the visibility of data fields but not their content. We thus are not sure whether a user subscribed to any newsfeed.

The fields gender, e-mail and birthday are obligatory to create a user profile on Facebook. Hence, every user profile encloses this data (not necessarily honest).  None of the other profile fields are filled by all users. The fields family, current city, relationship status, hometown, employer and school are filled with data by the majority of users. Only few FPW users uploaded skills and phone numbers to Facebook. Please note that we can only check whether data is included or not. We have no means to verify it. %Nevertheless, we assume most details to be true since misleading information, such as fake birthdays, potentially cause complications in case of sharing with friends.

\subsection{Visibility of User Profiles Fields}
\label{subsec:privacy_settings}

\begin{figure}[ht]
\centering
\includegraphics[width=0.477\textwidth]{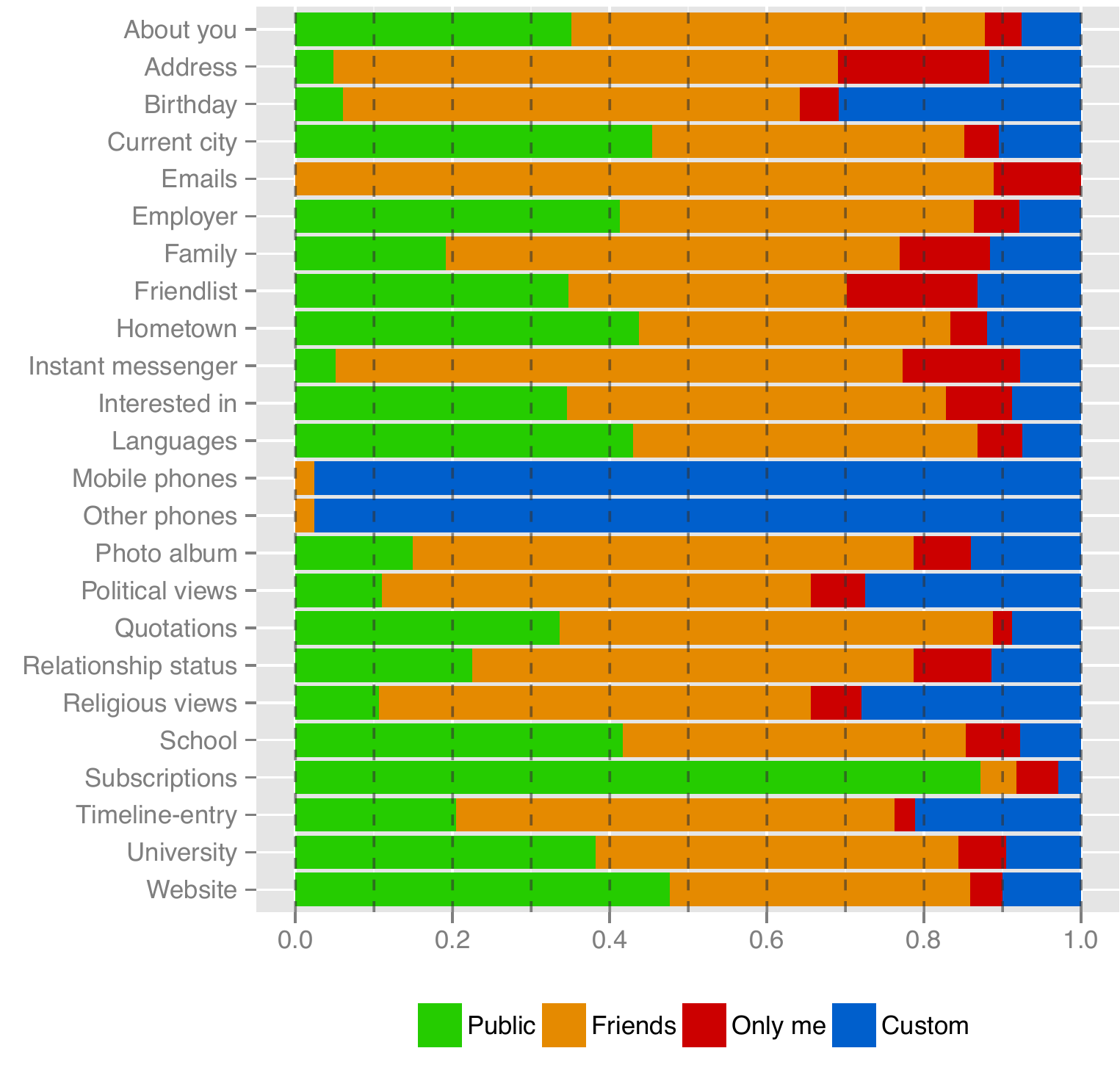}
\caption{Visibility of user profile fields}
\label{fig:privacy_settings_per_field}
\end{figure}

Figure \ref{fig:privacy_settings_per_field} shows the cumulated visibility of the profile fields of FPW users. The most popular setting is to share content items with all friends. The second most frequently used setting is to share items with the public. Sharing bits of information with only a subset of friends ('custom') or hiding them ('only me') is not very popular. 

More than one third of the users do not restrict access to the fields: current city, employer, friend list, hometown, languages, school and university. These profile fields may help attackers to collect sufficient information to deploy social engineering attacks. The friend list is especially dangerous to publish, since sharing the friend list helps attackers to traverse through the social graph using crawlers. Furthermore, inference attacks \cite{inference} are fostered by publishing the friend list. These kinds of attacks are based on the assumption that friends share similarities (e.g. similar age). An attacker can infer hidden profile attributes in case that friendship connections are known to the attacker and friends disclose the information of interest.

The custom setting is used for phone numbers in more than 95\% of those cases where this information is included into the user profile. More than a quarter of our study participants share the birthday, political views and religious views just with a subset of their friends. The fact that a non-negligible number of users use the setting 'only me' is remarkable. It makes sense that people disclose information in fields that are technically necessary (e.g. the friend list) in case that they do not want to share them with others. However, uploading other fields to Facebook without sharing it with anybody does not help to socialize with others. We assume fields with this visibility setting to be a result of increased privacy awareness. Previously visible informations seems to be hidden. 

\begin{figure}[ht!]
\centering
\includegraphics[width=0.477\textwidth]{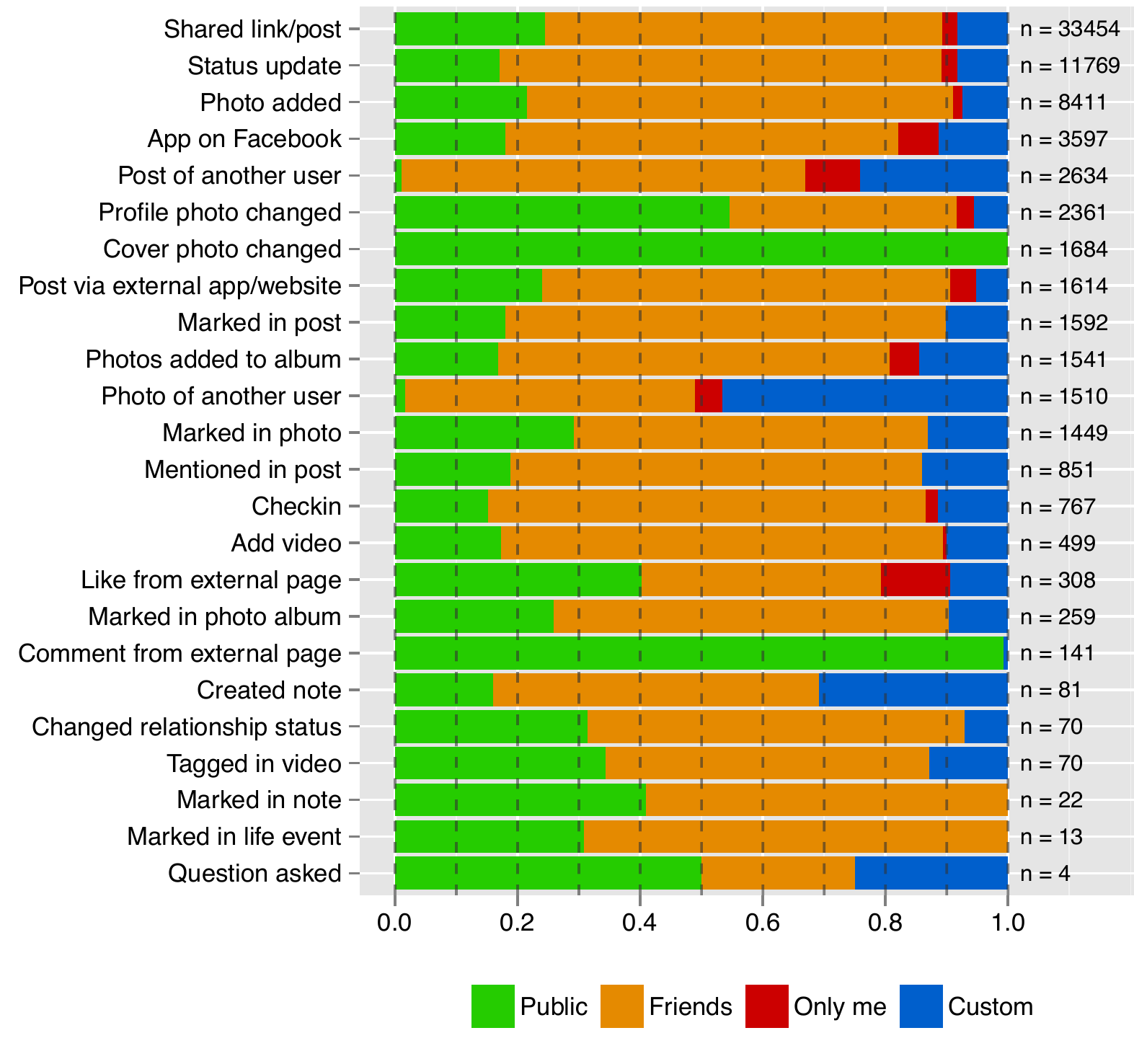}
\caption{Privacy settings of timeline entries}
\label{fig:timeline_entries_privacy_settings}
\end{figure}

% \subsection{Visibility of Timeline Entries}
% 
% Timeline entries are very popular. They can have 24 different types. We thus examine them in detail and dedicate this subsection. Figure \ref{fig:timeline_entries_privacy_settings} shows the visibility of all types of timeline entries. The main findings are that:

Timeline entries are similar to posts in a newsfeed and can have many different types. Figure \ref{fig:timeline_entries_privacy_settings} shows the visibility of all types of timeline entries. The main findings are that:

\begin{itemize}
 \item the setting 'friend' is even more dominant than in other parts of the profile
 \item less entries are visible to the public
 \item posts from external pages (e.g. commercial pages) and cover photo changes are always public
 \item the setting 'only me' is rarely used in general
 \item the most frequently hidden timeline entries are likes from external pages, posts from other users and posts from apps 
 \item photos of other users are often shared with only a subset of friends
\end{itemize}

\subsection{Privacy Impact of Simplified Audience Selection}
\label{subsection:impact_interface}

%The privacy impact of the plug-in can be measured by comparing the privacy settings which were found by the plug-in at the time of the first start of the plug-in with the settings after using the plug-in. 

Many Facebook users are unable to handle the privacy settings to meet their own sharing preferences \cite{liu2011analyzing,Madejski}. It is hence not sufficient to elaborate the actual privacy settings to study the sharing preferences. Since the color-coding based privacy setting interface is shown to drastically decrease mistakes in selecting the audience \cite{paul2012c4ps}, elaborating the impact of the FPW helps to understand the gap between sharing interests and actual privacy settings.

With the help of our plug-in, 22.31\% of the users change the visibility to a more restrictive setting, 19.55\% of the users prefer less restrictive settings and 5.44\% keep the average privacy by changing the visibility of different items equally to both directions. 52.14\% of the users do not change the profile visibility compared to the settings before installing our plug-in. 

The group of users who did not change any setting contains many inactive people with small user profiles as well as those who sent us feedback during the first session with activated FPW. All users who were not able to change any setting because of facing technical problems are also part of this group. In spite of not changing the settings, some users sent us feedback to state that the plug-in is very useful to check the settings with very little effort.

% \begin{figure}[ht]
% \centering
% %\includegraphics[width=1\textwidth]{grafiken/privacy_of_fields_changed.pdf}
% \includegraphics[width=0.489\textwidth]{grafiken/privacy_of_fields_changed.pdf}
% \caption{Fields changed be the users}
% \label{fig:fields_changed}
% \end{figure}

\begin{figure}[h]
\centering
\includegraphics[width=0.477\textwidth]{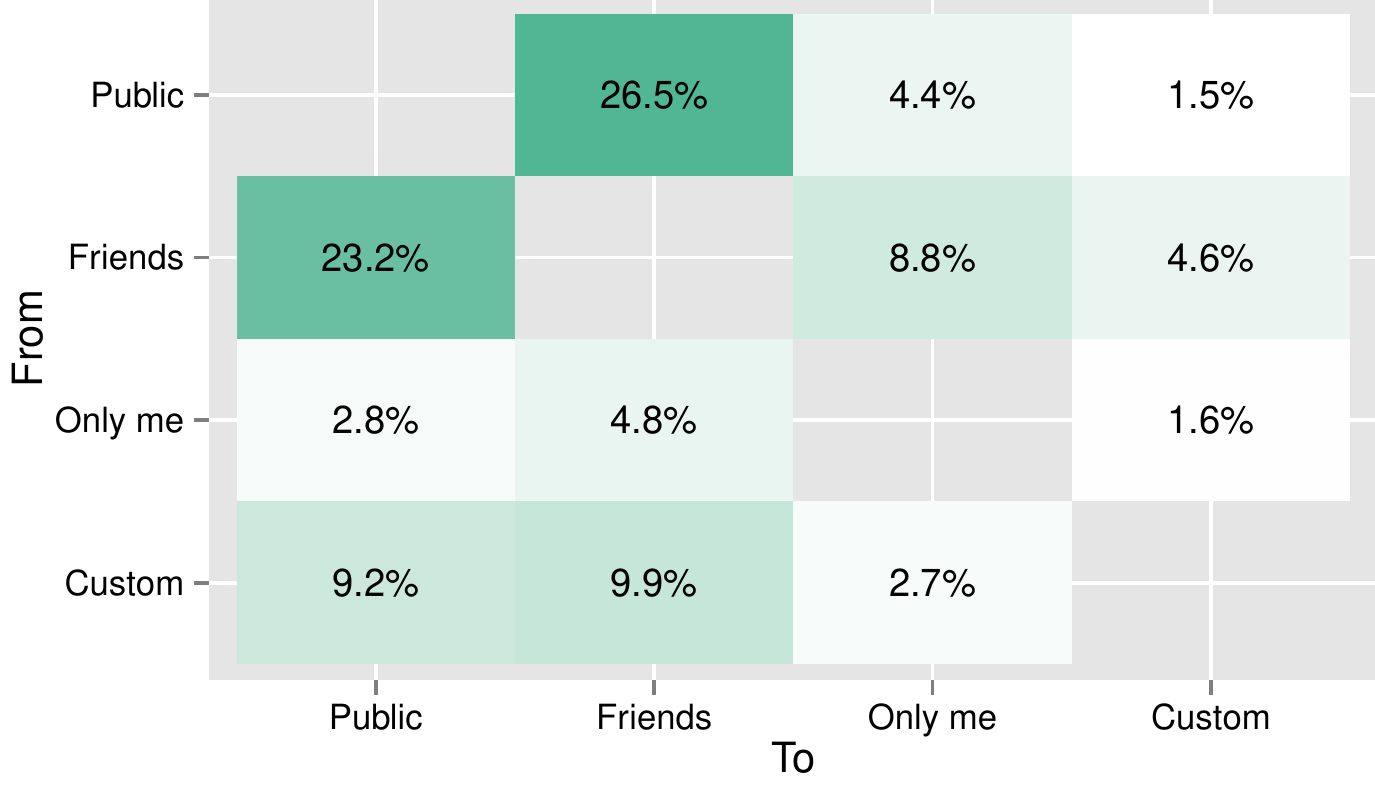}
\caption{Heat map of visibility levels reflecting visibility change actions, performed with the help of the new interface (from, to)}
\label{fig:fields_changed_matrix}
\end{figure}

In the remainder of this section, we focus on users who change the visibility of profile fields using the FPW. Figure \ref{fig:fields_changed_matrix} shows a heat map that illustrates change actions with respect to the visibility level before and after performing the actions. % with our plug-in. 
The most frequently performed action is to change the visibility from 'public' to 'friends'. The opposite change action is the second most frequently performed action. 

With the help of the FPW, users hide more information ('only me') from public or friends than providing access to content. Remarkable is that the custom visibility setting, which is explicitly supported by our interface, is more likely to be removed than being newly used. Many users seem not to be happy to distinguish among different groups of friends. They instead prefer to either publish content without restrictions or among all friends.

\begin{figure}[h]
\centering
\includegraphics[width=0.477\textwidth]{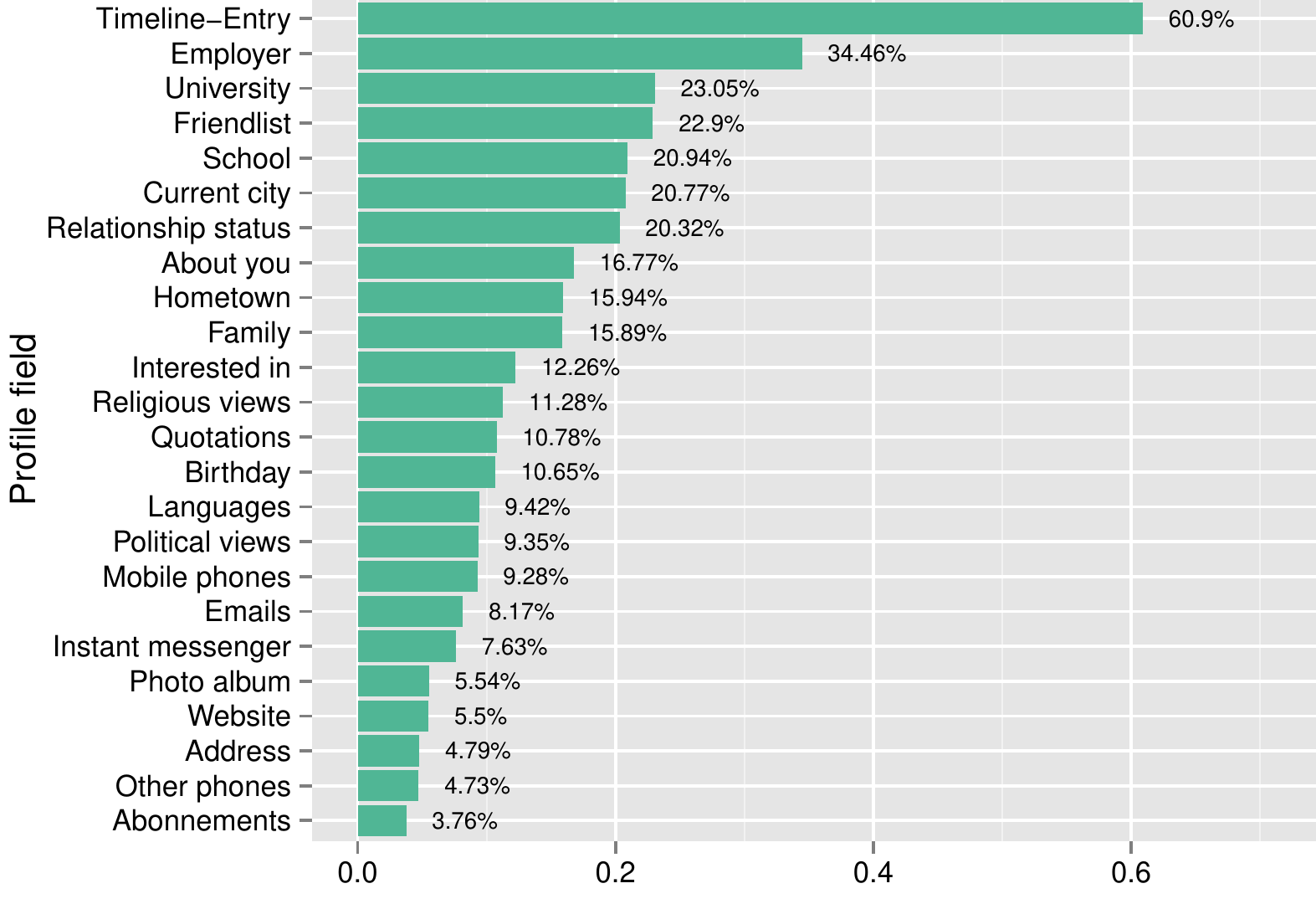}
% \caption{Percentage of filled profile fields which are changed using the FPW}
\caption{Percentage of users who changed the visibility of certain profile fields; only filled fields are mentioned}
\label{fig:fields_changed_field_number}
\end{figure}

Figure \ref{fig:fields_changed_field_number} depicts the exact percentage of items per profile field where users changed the visibility with the FPW. We only included those 2816 users whose privacy has finally been affected by the FPW. The highest demand for changes can be seen in the timeline entries. A user profiles in Facebook can enfold plenty of timeline entries but only a single entry in many other fields (e.g. birthday). The visibility of the employer has been changed by the second largest fraction of users, followed by the university and the friend list. %There is no discrepancy in the fact that 52.14\% of FPW users did not change anything and  a fraction 55.43\% of all timeline entry visibility settings which have been changed.

\begin{figure}[ht]
\centering
\includegraphics[width=0.477\textwidth]{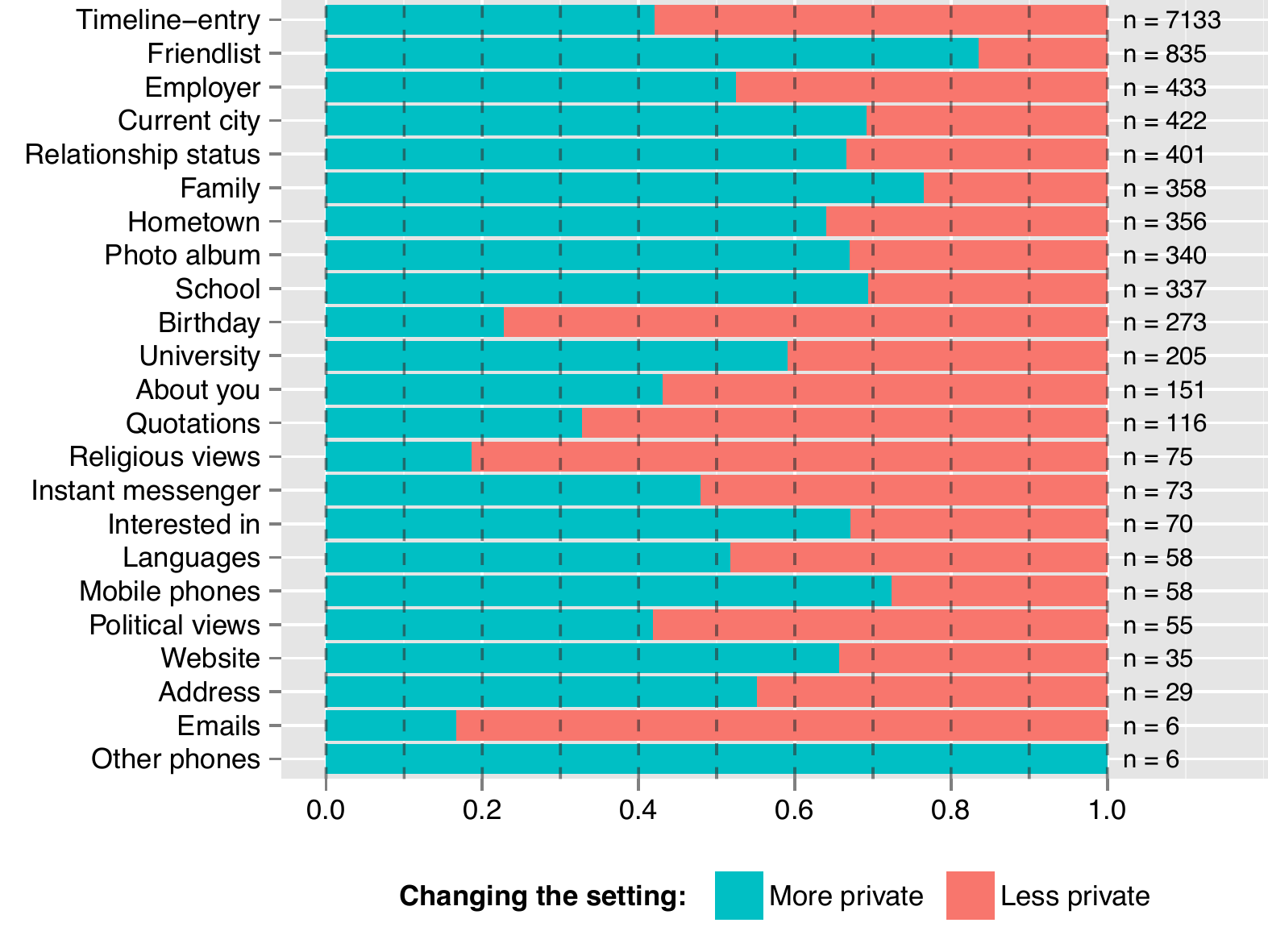}
\caption{Fraction of change actions with the help of the FPW towards more or less privacy per profile field}
\label{fig:fields_changed_field_tendency}
\end{figure}

The tendency of performed changes towards more or less privacy in different profile fields is shown in Figure \ref{fig:fields_changed_field_tendency}. Timeline entries, birthdays, about you, quotations, religious views, instant messagers, political views and e-mail addresses are those fields where more change actions towards less privacy have been performed. The rest of the profile fields are more private in average after using the FPW.

\subsection{Comparison with Facebook Standard Privacy Settings}

% In case that the CEOs of Facebook and Google would be right in saying that privacy is not important \cite{gogleprivacy,facebookprivacy}, users could save the efforts to change the default privacy settings or could simply choose to publish everything. We thus compare the user's audience choices with Facebook's default settings. 

Advocates of the concept 'privacy by default' argue that people do not tend to change the default settings. Following this argumentation, and taking the user's audience selection efforts into account, an interesting question is how the defaults should look like to be in line with the user's needs. We thus compare the default settings with the actual privacy settings.

\begin{figure}[ht]
\centering
\includegraphics[width=0.477\textwidth]{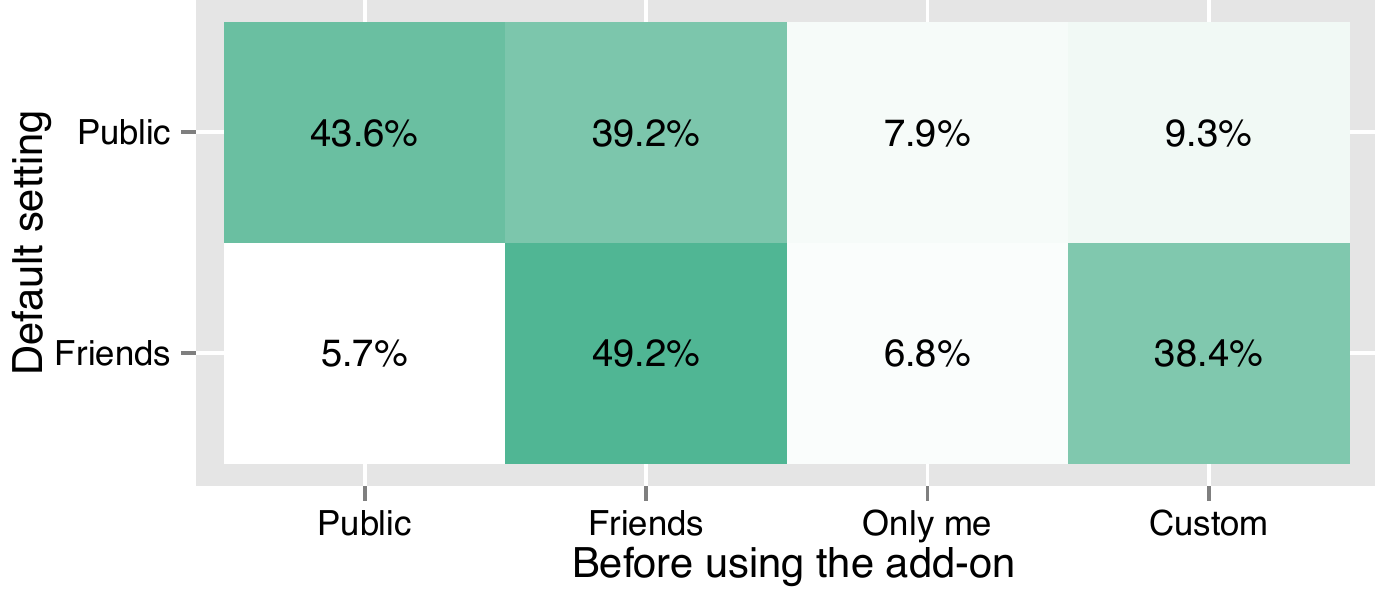}
% \caption{Comparison of Facebook standard visibility with the profile visibility before using the new interface}
\caption{Heat map that illustrates the privacy setting changes from Facebook standard (ordinate) to individual settings (abscissa) before using the new interface}
\label{fig:comparison__with_standard_before}
\end{figure}

The Facebook default settings consist of two visibility levels: public and friends. The heat map in Figure \ref{fig:comparison__with_standard_before} shows a comparison of the standard settings with the condition before applying the changes with the new interface:  43.6\% of all profile fields, which are shared with public according to the Facebook standard, are publicly accessible.  39.2\% of these public fields have been changed to be accessible only by friends. 49.2\% of the by default friend-visible profile fields are still friend-visible before using the FPW  and 38.4\% of profile of the latter are visible to just a subset of friends.

% Please note: these differences show the cumulated changes of those which have been performed with the Facebook standard interface and our new interface.

\begin{figure}[ht]
\centering
\includegraphics[width=0.477\textwidth]{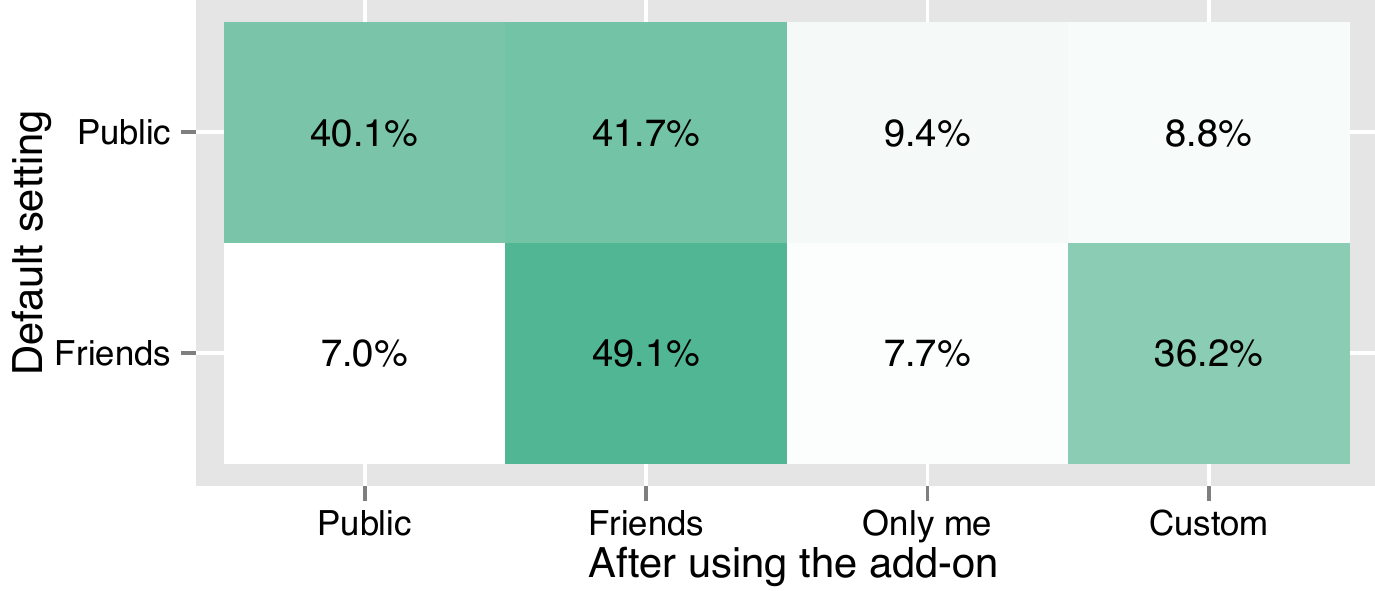}
\caption{Heat map that illustrates the privacy setting changes from Facebook standard (ordinate) to individual settings (abscissa) after using the new interface}
\label{fig:comparison__with_standard}
\end{figure}

Figure \ref{fig:comparison__with_standard} illustrates the comparison of standard settings with the situation after using the FPW. In spite of many users changing profile settings, the cumulated amount of visible content does not change dramatically. 21.05\% of the users used the plug-in to reduce the visibility of data objects in average by changing the standard settings. 10.44\% changed the standard settings to the opposite direction. Our evaluation shows that the visibility of profile fields is still conform with the standard settings in many cases. 40.56\% of the public fields are still unchanged after using the plug-in. That is also true for 49.93\% of the fields which are friend-visible by default.

\section{Country-Specific Privacy Evaluations}
\label{sec:country_specifics}

Since privacy preferences are depending on cultural backgrounds of users \cite{dataprivacy}, we detail the global evaluations by comparing the actual privacy settings as well as the impact of the FPW with respect to the user's country of origin. Due to space limitations, we abstain from including every single profile field and concentrate on the examples showing the strongest variations.

% \todo{
% \subsection{Sample Size Implications on Cross-Country Comparisons}

As a result of constraints in our dataset, the cross-country comparisons suffer from differences in sample sizes. We address this issue in the following evaluations by normalizing all data and comparing only fractions (proportions) and medians which are rather stable with respect to different sample sizes. Also, we only include samples which are big enough to be stable against outliers and only apply extremely conservative statistic testing. %and it is still bigger than studies using interviews or surveys
Since we used the same method for acquiring study participants in all countries, we assume a potential bias to equally occur amongst the considered countries. Hence, we assume the comparability of our samples from various countries to be valid. Germany is a special case since our university is well known and receives more attention and trust here.

% - smallest sample size is still 

% }

\subsection{Exposure and Visibility of Personal Data in Different Countries}
\label{subsec:privacysettings_countries}

\begin{figure}[ht]
\centering
\includegraphics[width=0.477\textwidth]{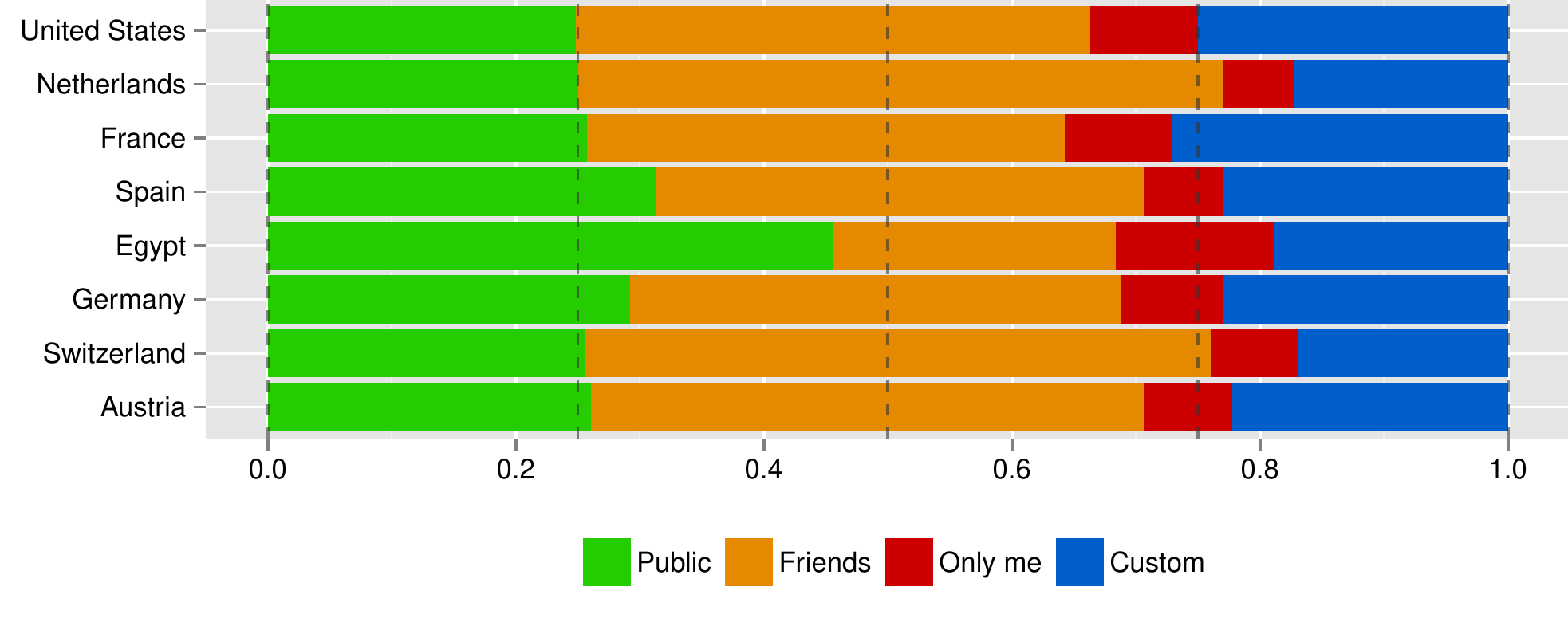}
\caption{Cumulated privacy settings in different countries; sample sizes can be found in Table 1}
\label{fig:privacy_per_country_summed}
\end{figure}

FPW users from different countries have different sharing interests. This can be shown by comparing both: the information which is enclosed into the user profiles (filled fields) as well as privacy settings. Figure \ref{fig:privacy_per_country_summed} shows the cumulated differences among the eight countries with feedback of more than 50 users. We cumulate all profile fields of all users in the respective country and compare the total proportions of content according to their visibility.

The most obvious result in our evaluation is that Egyptian users tend to share more information with the public than others. The latter also tend to hide the highest fraction of information (setting: 'only me') from anybody. Compared with the other seven countries, they tend to either publish content or not, rather than sharing with friends. We thus formulate the hypothesis that people in Egypt tend to use their Facebook profile as a tool to present themselves rather than to share content with their friends. Users from other Arabic countries seem to show a similar behavior, but the sample size is too small to provide meaningful results to include them into this paper.

French users include the highest fraction of content to their profiles which is visible for just a subset of their friends. FPW users from Germany and the USA show significant differences in hiding content from others (setting: 'only me'). Many other differences can be seen (Figure \ref{fig:privacy_per_country_summed}), but they are not significant according to our extremely strict criteria.

We tested the significance of country-specific differences by applying the Mann–Whitney U-test (with continuity correction) on four distinct datasets. We compared (country pair-wise on user granularity) the country-specific percentages of the user profile field visibility to be either 'public', 'only friends', 'only me' or 'custom'. The Benjamini \& Hochberg correction \cite{benjamini1995controlling} has been applied to adjust p-values for multiple comparisons (28 pairwise comparisons). %and we thus consider only results of p-values smaller than $0.05/7$ to be significant. This correction is the result of the 0.05 significance level and the fact that each country-specific dataset has been compared to seven others. Please note that this evaluation is extremely conservative and differences need to be very strong to be considered as significant. 
Table \ref{tab:country_sig_total} provides the results.

\begin{table}
\scriptsize
 \begin{center}
\begin{tabular}{rlllll}\hline
Country & Country & W & p-value & BH & Setting\\ \hline
Egypt & Austria & 3926 & 0.00010 & 0.00131  & Friends\\
Egypt & Switzerland & 1613 & 0.00015 & 0.00131 & Public \\
Egypt & Switzerland & 2380.5 & 0.00010 & 0.00131 & Friends \\
Egypt & France & 1974.5 & 0.00080 & 0.00378 & Public \\
Egypt & Netherlands & 1800.5 & 0.00039 & 0.00221 & Public \\
Egypt & Netherlands & 484 & 0.00018 & 0.00131 & Friends \\
Germany & USA & 29431 & 0.00779 & 0.02726 & Only me\\
France & Switzerland & 752.5 & 0.00533 & 0.02133 & Custom\\ \hline
 \end{tabular}
 \end{center}
\caption{Subset of significant results of the Pairwise Mann–Whitney U test of cumulated the data in Figure \ref{fig:privacy_per_country_summed}; W = test statistic; BH = Benjamini
\& Hochberg correction for multiple comparisons}
\label{tab:country_sig_total}
\end{table}

%  [1] "Country1: AT, Country2: EG, Friends:"
% W = 3926, p-value = 0.0001006
% 
% [1] "Country1: CH, Country2: EG, Public:"
% W = 1613, p-value = 0.0001558
% 
% [1] "Country1: CH, Country2: EG, Friends:"
% W = 2380.5, p-value = 0.000109
% 
% [1] "Country1: DE, Country2: US, Only me:"
% W = 29430.5, p-value = 0.00779
% 
% [1] "Country1: EG, Country2: FR, Public:"
% W = 1974.5, p-value = 0.0008093
% 
% [1] "Country1: EG, Country2: FR, Custom:"
% W = 1112.5, p-value = 0.006959
% 
% [1] "Country1: EG, Country2: NL, Public:"
% W = 1800.5, p-value = 0.0003944
% 
% [1] "Country1: EG, Country2: NL, Friends:"
% W = 484, p-value = 0.0001875
% 
% [1] "Country1: FR, Country2: CH, Custom:"
% W = 752.5, p-value = 0.005332

\begin{figure}[h!]
\centering
\includegraphics[width=0.477\textwidth]{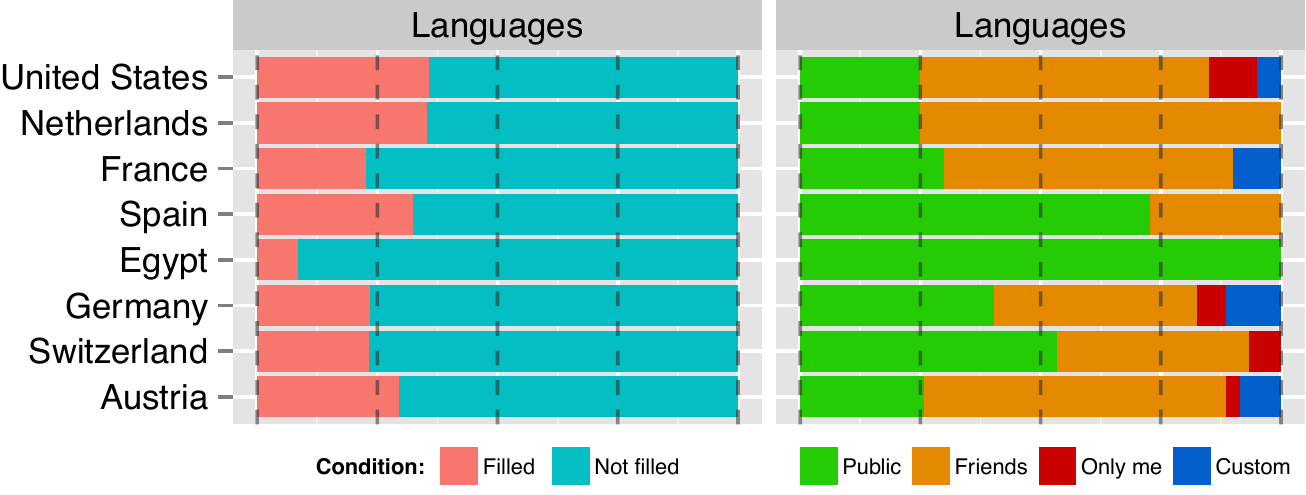}
\caption{Privacy settings of the field 'languages'}
\label{fig:field_languages}
\end{figure}

\begin{figure}[ht]
\centering
\includegraphics[width=0.377\textwidth]{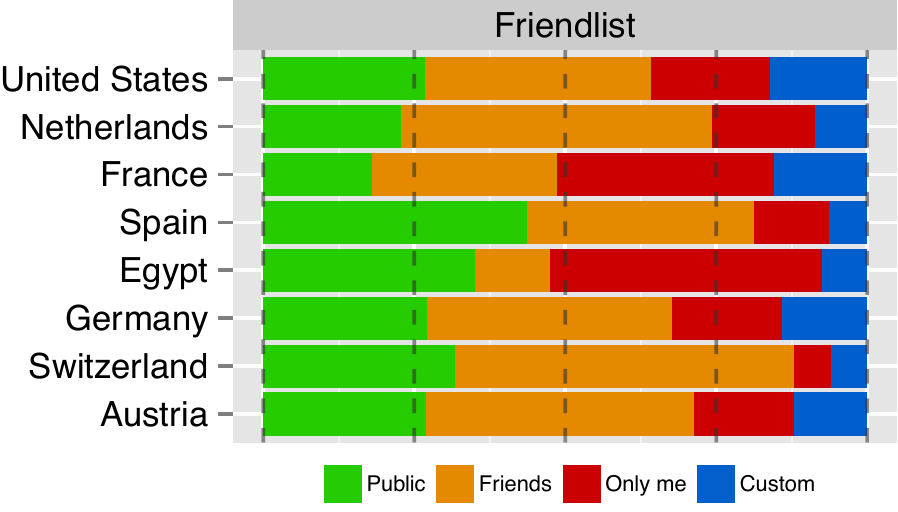}
\caption{Privacy settings of the field 'friend list'}
\label{fig:field_friendlist}
\end{figure}

Country-specific content sharing differences can be even stronger realized by comparing the visibility of certain profile fields in different countries. We thus choose a sample of seven fields to explain the differences in Figures \ref{fig:field_languages} %, \ref{fig:field_mobile_phones}, \ref{fig:field_hometown},\ref{fig:field_religious_views} and 
 till \ref{fig:field_friendlist}.

Evaluating the languages field (Figure \ref{fig:field_languages}), we realized that Egyptian users do only rarely include the languages into their profiles. However, in case they do, they share this information with the public. This is a very different behavior, compared to other countries. We would thus suspect Egyptians not to speak other languages very often but in case they do, they seem to be very proud of it. Spanish users do share the information about their languages significantly more often than users from USA and Austria. That is less significant but still valid for Swiss users, too.

\begin{table}
\scriptsize
\begin{center}
\begin{tabular}{rlllll}\hline
Country & Country & W & p-value & BH & Field\\ \hline
Egypt & Austria & 213.5 & 0.00064 & 0.01352 & Languages\\
Egypt & France & 10.5 & 0.00702 & 0.01776 & Languages \\
Egypt & Germany & 4613 & 0.00324 & 0.01469 & Languages \\
Egypt & Netherl. & 10.5 & 0.00248 & 0.01469 & Languages \\
Egypt & USA & 17.5 & 0.00154 & 0.0143 & Languages \\
Spain & Austria & 289 & 0.00539 & 0.0151 & Languages\\ 
Spain & USA & 53 & 0.00970 & 0.0209 & Languages\\
Egypt & Netherl. & 343.5 & 0.00097 & 0.01352 & Hometown \\
Egypt & Germany & 20119 & 0.00357 & 0.01469 & Religious V. \\
France & Egypt & 399.5 & 0.00407 & 0.01469 & Family \\
France & Germany & 22835 & 0.00761 & 0.01776 & Family \\
France & Netherl. & 495 & 0.00499 & 0.01508 & Family \\
France & Switzerl. & 316.5 & 0.00420 & 0.01469 & Family \\
 \hline \end{tabular}
 \end{center}
\caption{Subset of significant results of the Pairwise Mann–Whitney U test of non-cumulated data; W = test statistic; BH = Benjamini
\& Hochberg correction for multiple comparisons}
\label{tab:country_sig_languages}
\end{table}

\begin{figure}[h]
\centering
\includegraphics[width=0.477\textwidth]{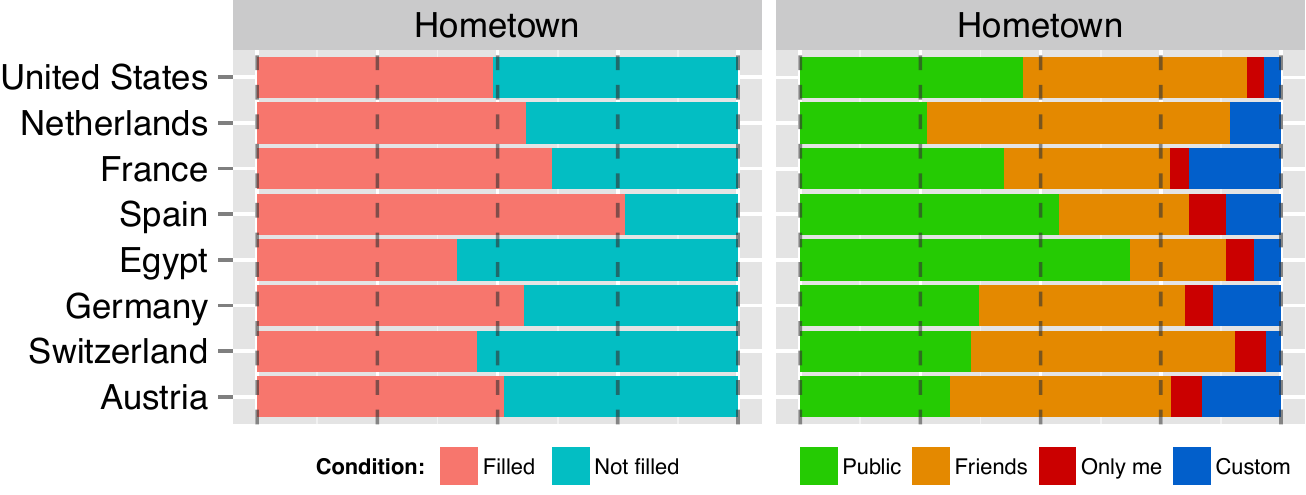}
\caption{Privacy settings of the field 'hometown'}
\label{fig:field_hometown}
\end{figure}

Another country-specific difference in sharing interest can be observed at the profile field 'Hometown' (Figure \ref{fig:field_hometown}). Egyptian FPW users share the name of the hometown with a significantly higher probability with the public than FPW users from the Netherlands. However, the highest fraction of users who added the hometown to the user profile is from Spain.  

\begin{figure}[ht]
\centering
\includegraphics[width=0.477\textwidth]{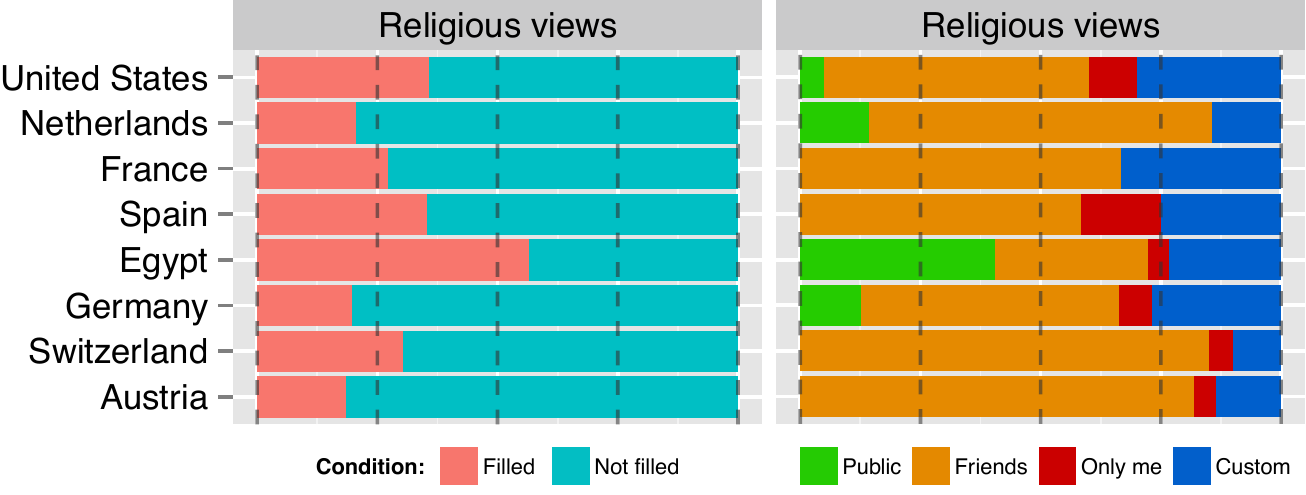}
\caption{Privacy settings of the field 'religious views'}
\label{fig:field_religious_views}
\end{figure}

The religious views (Figure \ref{fig:field_religious_views}) are less likely to be included in the Facebook profile of the FPW users than e.g. the hometown or the family status. Only among Egyptian users, a majority of people can be observed to add the religious views to the user profile in Facebook. Furthermore, the Egyptians form the group that publishes this information with the highest likelihood. This observation can be used to found the hypothesis that religious views and their public commitments are more important in Egypt than in the other countries that we consider in this paper.

\begin{figure}[ht]
\centering
\includegraphics[width=0.477\textwidth]{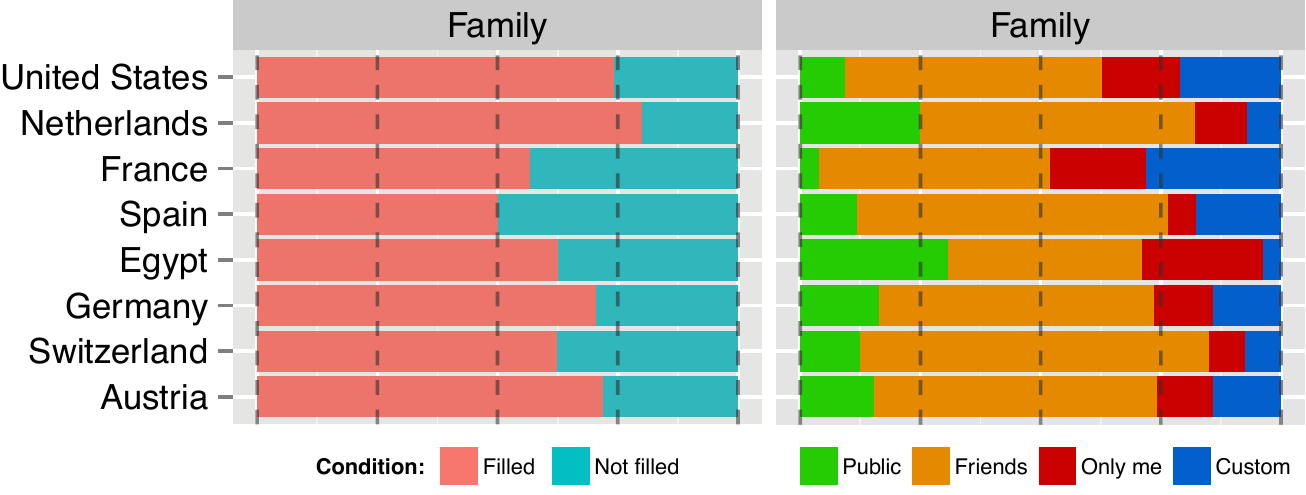}
\caption{Privacy settings of the field 'family'}
\label{fig:field_family}
\end{figure}

Information about the family status (Figure \ref{fig:field_family}) is very likely to be included into the profiles. The overwhelming fraction of users prefer to share this information only with friends. 
In comparison to others, French users tend to restrict access to this profile field. Remarkable is that this is the field which is hidden by the largest fraction of people.

\begin{figure}[ht]
\centering
\includegraphics[width=0.477\textwidth]{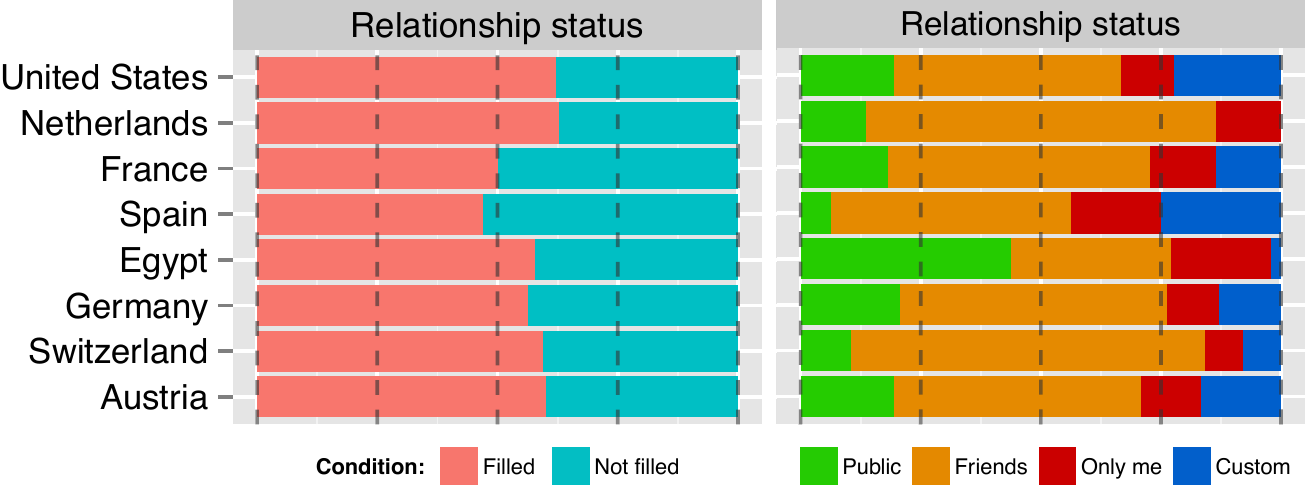}
\caption{Privacy settings of the field 'relationship status'}
\label{fig:field_relationship_status}
\end{figure}

Comparing the visibility of the relationship status of Spanish and Egypt FPW users (Figure \ref{fig:field_relationship_status}) is very interesting. Spanish FPW users are the subset with the lowest probability of filling and publishing the field 'relationship status'. With the highest probability compared to others, they share this information with only a selected subset of friends. In contrast, nearly half of the Egyptians publish their relationship status. At the same time, they are also the subset of FPW user with the highest likelihood to hide this bit of information.

The friend list (Figure \ref{fig:field_friendlist}) is the sole profile field in this evaluation which exists in every user profile without being empty. Users do not have the choice to upload a friend list or not: it is created automatically by adding friends. In case that users prefer not to share this information, their only chance is to hide the list by choosing the visibility setting 'only me'. Accordingly, the latter setting is very popular. This is especially true for the subset of Egypt FPW users. 

% \begin{figure*}[ht]
% \centering
% \includegraphics[width=0.89\textwidth]{grafiken/filled_per_country_part.pdf}
% \caption{Selected Profile Fields filled by Users from Different Countries; Fields with the biggest differences}
% \label{fig:fields_filled_per_country_part}
% \end{figure*}

% 
% \begin{figure*}[ht]
% \centering
% \includegraphics[width=0.89\textwidth]{grafiken/privacy_per_country.pdf}
% \caption{Fields changed be the users}
% \label{fig:privacy_per_country}
% \end{figure*}

\subsection{Country-Specific Changes of Privacy Settings}

In Section \ref{subsec:privacysettings_countries}, we elaborated the privacy settings in different countries and distinguished between different fields. The main finding was that users from different countries share different information with their friends or the public. Since all users are faced with the same default privacy settings while having different sharing desires, the necessity of changing the visibility settings to meet the own sharing desires thus also differs. In this section, we elaborate the change actions which have been performed with the help of the FPW with respect to the user's country of origin.

In the remainder of this section, we distinguish among four subsets of users. The first subset, denoted \emph{only less private}, consists of users who only changed visibility settings towards a higher visibility, e.g. from 'friends' to 'public' or from 'only me' to 'friends'.
The second subset consists of users who changed the visibility \emph{less private}. That means the users perform changes in both directions but those changes which grant more access to profile fields prevail the others. Accordingly, we denote the third and the fourth subset \emph{more private} and \emph{only more private} were the third subset consists of users who mainly changed to a more private setting and the fourth subset of users who only restricted access to profile fields. 

We ignored two subsets which could be built when following the previous logic: those users who did not change anything and those users who changed the privacy settings equally to both direction. The latter have been ignored since the subset contains many users who only tried our new interface and changed one field in both directions. The subset of users who did not change anything can hardly be evaluated since this subset contains those users who faced technical problems, thus unable to perform changes. 

\begin{table*}
\small
 \begin{center}
\begin{tabular}{p{2.7cm}|p{1cm}p{1cm}|p{1cm}p{1cm}|p{1cm}p{1cm}|p{1cm}p{1cm}|ll|l} \hline \hline \smallskip
\textbf{Cluster} & \multicolumn{2}{c}{\textbf{Friends}}&  \multicolumn{2}{c}{\textbf{Likes}}& \multicolumn{2}{c}{\textbf{Photos}}& \multicolumn{2}{c}{\textbf{Map entries}}&  \multicolumn{2}{c}{\textbf{Notes}}& \\  %\cline{2-11}

%         &
% 	    &
        & Mean & Median & Mean & Median & Mean & Median & Mean & Median & Mean & Median &\\
       
\hline

Only more private & 171.48 & 112 	&  106.03 & 39 & 22.34 & 3 & 38.69 & 4  & 0.48  & 0 \\
More private	  & 163.64 & 107		&  131.69 & 49 & 19.16 & 3 & 46.95 & 4  & 19.87 & 0 \\
Less private	  & 177.11 & 88		&  142.58 & 49 & 28.81 & 2 & 49.87 & 7  & 1.86  & 0 \\
Only less private & 186.49 & 91		&  125.69 & 55 & 22.69 & 4 & 98.32 & 11 & 0.71  & 0 \\

%  \hline 
 \end{tabular}
 \end{center}
\caption{Profile statistic comparison with respect to change direction clusters}
\label{tab:cluster_group}
\end{table*}

%Die Werte in der Tabelle beziehen sich nur auf die 4 Cluster
% \begin{table*}
%  \begin{center}
% \begin{tabular}{p{2cm}|p{1cm}p{1cm}p{1cm}p{1cm}p{1cm}p{1cm}p{1cm}p{1cm}lll} \hline \hline \smallskip
% \textbf{Country} & \multicolumn{2}{c}{\textbf{Friends}}&  \multicolumn{2}{c}{\textbf{Likes}}& \multicolumn{2}{c}{\textbf{Photos}}& \multicolumn{2}{c}{\textbf{Map entries}}&  \multicolumn{2}{c}{\textbf{Notes}}& \\  %\cline{2-11}
%         
% 
% %         &
% % 	    &
%         & Mean & Median & Mean & Median & Mean & Median & Mean & Median & Mean & Median &\\
%        
% \hline
% United States 	& 218.7  &  92 &  92.21 &  64 &  46.79 & 11 &  25.3  &  5 &  1.91 & 0 \\
% Netherlands 	& 113.45 &  91 &  32.76 &   8 &  18.14 &  8 &  51.69 & 24 &  0.03 & 0 \\
% France 		& 285.04 &  99 & 142.68 &  36 &  35.93 &  7 &  43.5  &  3 & 21.32 & 0 \\
% Spain 		& 159.88 & 132 & 143.5  &  27 & 106.81 & 61 & 114.06 &  9 &  3.41 & 0 \\
% Egypt 		& 617.0  & 281 & 271.83 & 190 & 126.0  & 24 &  57.0  &  7 & 17.17 & 8 \\
% Germany 	& 154.48 &  92 & 105.87 &  37 &  15.56 &  2 &  47.88 &  4 &  0.48 & 0 \\
% Switzerland 	& 245.76 & 173 & 204.07 &  38 &  31.54 &  7 &  82.87 &  9 & 70.37 & 0 \\
% Austria 	& 278.79 & 188 & 174.2  &  96 &  36.12 & 16 & 121.93 &  7 &  1.69 & 0 \\
% 
% %  \hline 
%  \end{tabular}
%  \end{center}
% \caption{Profile statistic comparison with respect to countries}
% \label{tab:cluster_country}
% \end{table*}

\begin{table*}
\small
 \begin{center}
\begin{tabular}{p{2cm}|p{1cm}p{1cm}|p{1cm}p{1cm}|p{1cm}p{1cm}|p{1cm}p{1cm}|ll|l} \hline \hline \smallskip
\textbf{Country} & \multicolumn{2}{c}{\textbf{Friends}}&  \multicolumn{2}{c}{\textbf{Likes}}& \multicolumn{2}{c}{\textbf{Photos}}& \multicolumn{2}{c}{\textbf{Map entries}}&  \multicolumn{2}{c}{\textbf{Notes}}& \\  %\cline{2-11}

%         &
% 	    &
        & Mean & Median & Mean & Median & Mean & Median & Mean & Median & Mean & Median &\\
       
\hline
United States 	& 201.41 & 115 & 156.47 &  73 &  45.64 &  8 &  24.68 &  4 &  1.93 & 0 \\
Netherlands 	& 111.63 &  90 &  33.50 &  10 &  17.70 &  8 &  47.73 & 21 &  0.03 & 0 \\
France 		& 267.81 & 112 & 129.22 &  44 &  47.78 & 10 &  82.63 &  5 & 30.86 & 0 \\
Spain 		& 148.78 & 117 & 127.97 &  23 &  98.37 & 60 & 102.05 &  7 &  2.94 & 0 \\
Egypt 		& 331.86 & 150 & 455.70 & 181 &  61.86 & 17 &  25.29 &  2 &  7.93 & 8 \\
Germany 	& 154.54 &  93 & 102.22 &  35 &  15.38 &  2 &  45.61 &  4 &  0.52 & 0 \\
Switzerland 	& 225.52 & 131 & 185.41 &  40 &  28.30 &  6 &  67.48 &  9 & 50.58 & 0 \\
Austria 	& 275.83 & 193 & 171.46 &  96 &  36.55 & 16 & 108.91 &  5 &  1.60 & 0 \\

%  \hline 
 \end{tabular}
 \end{center}
\caption{Profile statistic comparison with respect to countries}
\label{tab:cluster_country}
\end{table*}

% maybe some ANOVA would be useful to quantify the effects

\begin{figure}[h!]
\centering
\includegraphics[width=0.477\textwidth]{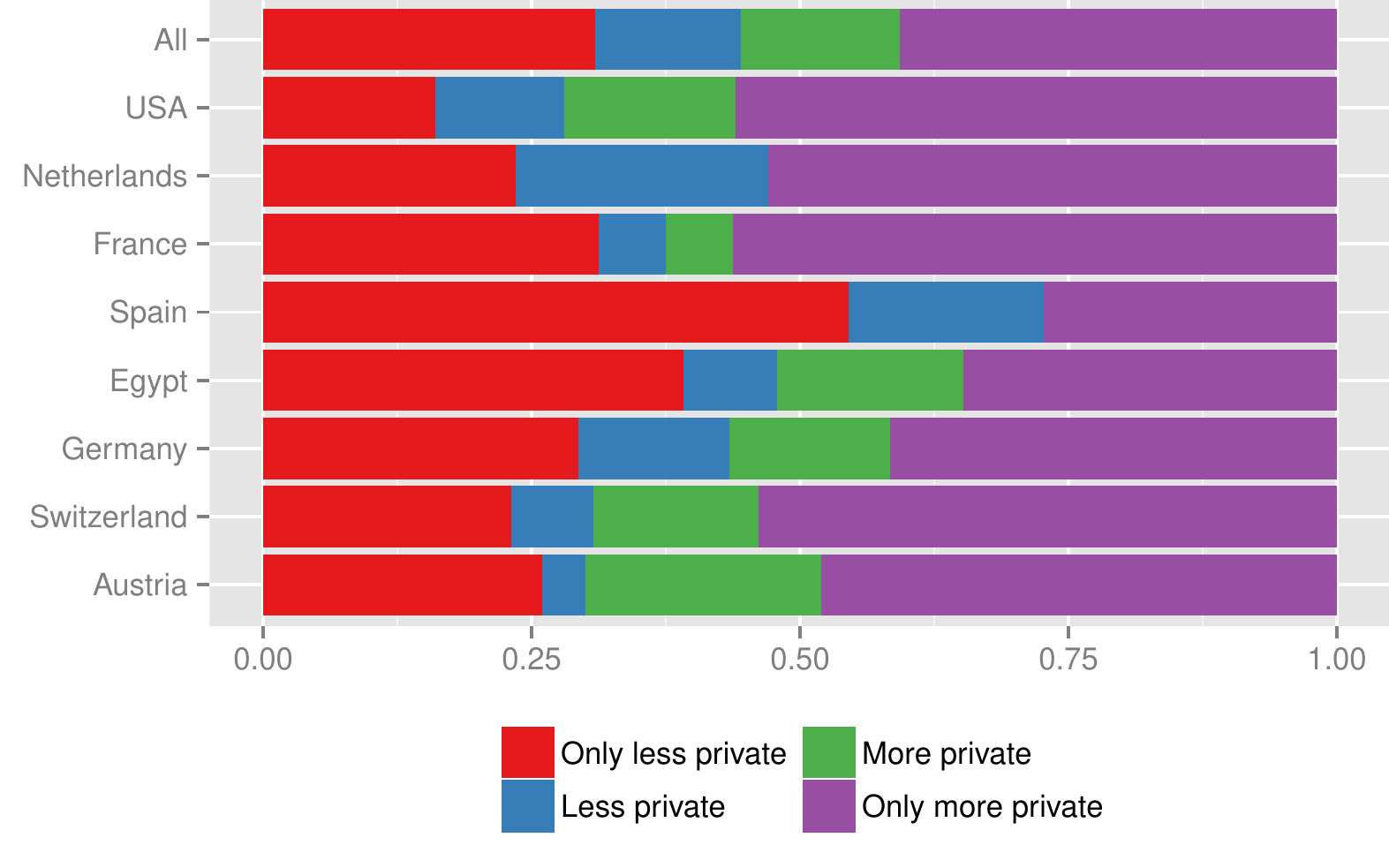}
\caption{Fractions of users grouped by change directions of actions with FPW}
\label{fig:change_country}
\end{figure}

Figure \ref{fig:change_country} shows the distribution of the four clusters in our top eight countries. The relative cluster sizes are different amongst the mentioned countries and the majority of the FPW users changes the visibility of profile fields towards one of the two possible directions. Surprisingly, in spite of advertising the FPW as a tool to increase the privacy, the fraction of users who only used the FPW to change the privacy settings to less private settings is relatively high (30.92\% of the sum of the four clusters). In Spain, the latter is even higher than 50\%. In total, the FPW caused less information to be accessible.

Comparing the privacy settings in Figure \ref{fig:privacy_per_country_summed} and the change actions in Figure \ref{fig:change_country} draws a homogeneous picture: The two countries with the least conservative settings are those with the highest fraction of users in the cluster \emph{only less private}. Switzerland and the Netherlands are at the opposite of the range in both illustrations.

\section{Change Direction Clusters}
\label{sec:clusteranalysis}

The clear distinction of clusters in Figure \ref{fig:change_country} inspired us to evaluate the differences in the user profiles to examine implications of privacy preferences on profile properties. In Tables \ref{tab:cluster_group} and \ref{tab:cluster_country}, we compare the mean and median of the countable profile properties 'Friends', 'Likes', 'Photos', 'Map Entries' and 'Notes' with respect to clusters and countries.

Users in the cluster \emph{only more private} have more friends (median) than others but less likes and less map entries. Users in the cluster \emph{more private} %who changed  the privacy settings towards both dire
still have more friends than those who used the FPW to increase the visibility of profile fields. Also notable is that users in the cluster \emph{only less private} do not mind to tell Facebook their location by having more map entries. Notes are not very popular amongst our set of users. The mean of 19.87 in the \emph{more private} cluster is a result of a fringe group of users having plenty of notes. 
% The median is still zero.

Table \ref{tab:cluster_country} shows the mean and the median of the same set of countable profile properties as they can be found in Table \ref{tab:cluster_group}. Obvious differences among country clusters are that Egyptian FPW users who sent us feedback have more friends and more likes than all others. The cluster of Dutch FPW users is the opposite extreme, having 18 times less likes (median) than Egyptian cluster. The Spanish users share 60 pictures, the German 2 (median).

% Our set of Egyptian FPW users have 150 friends (median) compared with Germans (93) or Dutch FPW users do. While Austria and Switzerland are similar in the number of friends, Germans do have only roughly half as many. Also very interesting are the differences in the number of likes. While our Egyptian FPW users have a median of 190 likes, the Dutch have 8 and the Spanish have 27. The median number of photos exhibits extreme differences: Germans share 2 pictures, Spanish FPW users 61.
% 
% When evaluating Table \ref{tab:cluster_country} row by row, some country-specific patterns can be discovered. Our Dutch users e.g. have the smallest number of friends and likes, a small number of photos and the largest median of map entries. They share and like just little content but seem to be very unconcerned to share their location. The opposite is true for the French FPW users. Compared with Dutch, the median of three map entries is eight times smaller and the median number of likes is 4.5 times bigger.

% The culture of Facebook usage seems to be very different

Suddenly, comparing the differences amongst our four change direction clusters in Table \ref{tab:cluster_group} exhibits notably smaller differences than comparing user profile differences amongst users from different countries in Table \ref{tab:cluster_country}. All values in Table \ref{tab:cluster_group} are very close to the values in the line 'Germany' in Table \ref{tab:cluster_country}. The reason is that the majority of the FPW users in this study is German. It underlines the influence factor \emph{country of origin} to dominate the \emph{change direction}.

% \todo{
% \section{Result Generalization}
% 
% 
% - broad discussion on the privacy topic: huge media attention from radio stations and newspapers -> privacy is assumed to be interesting topic for average recipients -> however the effort of installing FPW might not be taken by the privacy indifferent people
% - advertised as a tool to improve privacy -> however, a very big fraction of users disclose more information with the tool -> strong hint that it is not used by privacy freaks \\
% - does not meet the average user but is the best study in the wild
% }

\section{Summary and Conclusion}
\label{sec:conclusion}

In this paper, we presented the first large-scale study about content sharing and privacy preferences of Facebook users with special focus on country-specific characteristics. It is based on 9,292 feedbacks from 4,182 users in 102 countries. Our sample is neither complete nor a result of a random sampling process (Section \ref{sec:data_description}). Yet, the huge media attention from radio stations and daily newspapers, which address ordinary people, shows that the FPW was assumed to be interesting for their recipients. Furthermore, the fact that a very big fraction of users discloses more information instead of hiding it with the FPW is a strong evidence that it is not used by a fringe group of privacy savvy people. 

In contrast to related work in the field of privacy preferences, we %neither ask users in surveys nor crawl publicly-available data to evaluate users' privacy preferences. Instead, we 
collect our data on the users' clients and evaluate the behavior from real users who perform audience selection on their own user profiles for their own reasons. % on their own devices. 
However, even the evaluation of the actual privacy settings is only a rough estimation of the sharing preferences that suffers from two imprecisions: (i) Many users are unable to properly choose their audience with Facebook's privacy setting interface, and (ii) the sharing preferences exhibit a vast diversity depending on the user's country of origin. 

To overcome those imprecisions, we evaluated changes that have been made using color-coding based privacy controls. In a previous study, the latter have been demonstrated to be usable, intuitive and effective to drastically reduce errors and efforts in selecting the audience \cite{paul2012c4ps}. 

We further elaborated the country-specific differences in both the privacy settings as well as the privacy change actions. Additionally, a cluster analysis highlights the relation between the impact of the FPW on users' audience selection decisions and  their countable profile properties.

When creating an account in Facebook, it is obligatory to reveal information about gender, e-mail and birthday while creating an account on Facebook. However, our results indicate that the majority of FPW users sufficiently trusts Facebook to confide personal information such as family status, current city, hometown, employer and school. Contrariwise, only a minority of FPW users includes information on skills, addresses or political views into their profiles.

The most popular audience selection strategy is to allow all friends to access a certain bit of information, followed by publishing it and disclosing it to only a subset of friends. The setting 'only me' is the least popular setting. Beside unpopular features such as subscriptions and websites, the current city, the hometown, languages and the employees are the most frequently published bits. Only very few FPW users publish their e-mail address, instant messenger ID and their birthday, but the majority shares these bits with their friends. %Phone numbers (both mobile and others), being rarely shared on Facebook, are only disclosed to a subset of friends by the overwhelming majority of users. 
The friend list is a divisive issue amongst users to decide about its audience. Being published by more than one third of all FPW users, the friend list is the profile field that the second largest fraction of users is hiding (setting 'only me').

Introducing the comprehensible color-coding interface of the FPW impacts the audience selection of users. %In total, 47.3\% of the users changed changed their privacy settings with the help of our plug-in. 
In spite of the FPW being advertised as a privacy tool, users disclose selected bits of information to the public and to the complete set of friends. Users mainly change the privacy settings for timeline entries, the friend list and the profile field 'employer'. While the visibility of the timeline entries and the field employer are roughly equally switched to more and less restrictive privacy settings, the friend list setting was preferred to be more restrictive by 83\% of our participants. The total amount of content which is visible to Facebook users does not dramatically decrease after introducing a comprehensible visualization of privacy controls, but the composition of the visible content changes. This indicates that the usability of Facebook's privacy setting interface can be improved by using color codings. %The comparison of the actual privacy settings in the user profiles with the Facebook standard settings indicate that our users have a clear understanding about what to publish or not. Only 4.57\% of the FPW users entirely accept the default privacy settings in Facebook. 

Which information is uploaded to Facebook as well as which information is shared with whom is strongly depending on the user's country of origin. A perspicuous example is that less than 22\% of the German FPW users shared their religious views on Facebook while the majority of Egyptian FPW users included their religious views into their user profiles. The visibility is chosen accordingly. Thus, global default privacy settings cannot meet the sharing interests of all users since the sharing interests show country-specific as well as person-specific differences. 

Authors of alternative OSN architectures argue that fine-grained access control is an important feature to improve privacy in OSNs \cite{Jahid2011a,simpson2008need,carminati2009semantic}. However, our FPW users tend to remove group settings and individual access rules to achieve a lower complexity of access rules. %Even those Facebook users who invested the effort to download and install the privacy tool FPW preferred simple privacy settings. %That may be a hint that a very fine-grained access control mechanism is not a strong competitive differentiation criterion in the field of OSN.
We construe this fact to express user's favor for simplicity and thus encourage privacy interface designers to focus on simplicity rather than on a rich set of functionality.

\section{Acknowledgements}
This work has been co-funded by the German Research Foundation (DFG) in the Collaborative Research Center (SFB) 1053 'MAKI – Multi-Mechanisms-Adaptation for the Future Internet.

% \tp{
% 
% Unterschiedliche Nutzungspraktiken in verschiedenen Laendern -> Arbeiten die auf Usern in einem einzigem Land basieren haben einen limited scope!}

% \newpage

\bibliographystyle{plain}
\bibliography{bib/Journal,bib/p2psn}

%----------------------------------------------------------------------

\end{document}